\documentclass[epj,nopacs]{svjour}
\usepackage{epsfig,array}
\usepackage{multirow,color}
\usepackage{epsf,graphicx}
\usepackage{times,latexsym,euscript,amstext,amssymb,amsbsy,amsmath,amsfonts}
\usepackage{mdwlist}\usepackage{epsfig}
\usepackage{bm,rotating}
\usepackage{slashbox}
\newcommand{\be}{\begin{eqnarray}}
\newcommand{\ee}{\end{eqnarray}}
\newcommand{\bc}{\begin{center}}
\newcommand{\ec}{\end{center}}
\newcommand{\bea}{\begin{eqnarray}}
\newcommand{\eea}{\end{eqnarray}}

\newcommand{\beq}{\begin{equation}}
\newcommand{\eeq}{\end{equation}}

\def\fun#1#2{\lower3.6pt\vbox{\baselineskip0pt\lineskip.9pt
\ialign{$\mathsurround=0pt#1\hfil##\hfil$\crcr#2\crcr\sim\crcr}}}

\begin{document}

\title{\boldmath Pion- and photo-induced transition amplitudes to $\Lambda
K$, $\Sigma K$, and $N\eta$ }
\titlerunning{Pion- and photo-induced transition amplitudes to $\Lambda
K$, $\Sigma K$, and $N\eta$ }
\author{
A.V.~Anisovich$\,^{1,2}$, R. Beck$\,^1$, E.~Klempt$\,^1$,
V.A.~Nikonov$\,^{1,2}$, A.V.~Sarantsev$\,^{1,2}$,  and
U.~Thoma$\,^{1}$ }
\authorrunning{A.V.~Anisovich \it et al.}
\institute{$^1\,$Helmholtz-Institut f\"ur Strahlen- und Kernphysik,
Universit\"at Bonn, Germany\\
$^2\,$Petersburg Nuclear Physics Institute, Gatchina, Russia}

\date{Received: \today / Revised version:}

\abstract{Pion and photo-induced inelastic reactions off protons are
studied in a multichannel partial-wave analysis. Properties of
nucleon and $\Delta$ resonances are derived and compared to previous
analyses. Amplitudes are shown for transitions to $N\eta$, $\Lambda
K$, and $\Sigma K$ final states.
 \vspace{1mm}   \\
 {\it PACS:
11.80.Et, 11.80.Gw, 13.30.-a, 13.30.Ce, 13.30.Eg, 13.60.Le
 14.20.Gk}}

\maketitle

\section{Introduction}
The spectrum of baryon resonances, their masses, widths, and
photo-couplings \cite{Klempt:2009pi} provide a wealth of information
to test theories modeling QCD \cite{Isgur:1977ef,Isgur:1978xj,%
Capstick:1986bm,Loring:2001kx,Glozman:1997ag,Brodsky:2006uqa,%
Forkel:2008un} or, eventually, calculations of these quantities on
the lattice \cite{Durr:2008zz,Dudek:2009qf,Bulava:2010yg,%
Engel:2010my,Edwards:2011jj}. The Particle Data Group
\cite{Nakamura:2010zzi} summarizes regularly the experimental
information coming from a variety of different experiments. Some of
the information is already highly condensed: information on the
excitation spectrum of the nucleon stems from experiments on $\pi N$
elastic scattering, including experiments on $\pi^-p\to n\pi^+$
charge~ex\-change, and from experiments in which the target nucleon
is polarized, and from experiments in which the recoil polarization
of the scattered nucleon is measured in a secondary reaction. All
this information is exploited to determine energy-independent
partial wave amplitudes. In a second step, these amplitudes are
analyzed to extract resonant contributions \cite{Hohler:1979yr,%
Hohler:1993xq,Cutkosky:1980rh,Arndt:2006bf}. Partial decay widths
stem from dedicated experiments studying, e.g. pion induced hyperon
production or photoproduction of the known resonan\-ces.

Unfortunately, the experimental information is not really sufficient
to determine energy-independent partial wave amplitudes from $\pi N$
elastic scattering. Data on the nucleon recoil polarization do not
cover the full energy and angular range. They help to differentiate
between different partial wave solutions but they cannot be used to
construct the amplitudes from the data. Without this information,
only the absolute values of the spin-flip and spin non-flip
amplitudes can be determined but not their relative phases. The
recoil polarization is accessible easily in case of production of
hyperons which reveal their polarization by their asymmetric $N\pi$
decays. Unfortunately, pion-induced hyperon production experiments
with polarized targets have been carried only at a few momenta.
Hence again, the data do not suffice to reconstruct the amplitudes
in single energy bins. Even if the amplitudes are enforced to obey
some continuity criteria from energy bin to energy bin, the problem
does not converge to a unique solution. Dispersion relations -- for
fixed four-momentum transfer, for fixed cms-angles, and for fixed
energy -- can be used to relate the real and imaginary part of the
scattering amplitude and can thus, in principle, be exploited to
overcome the lack of experimental data. Further input can be derived
from a comparison of the total and differential cross section for
$\pi^{\pm}p$ and the Coulomb interference and from $\bar pp\to 2\pi$
exploiting backward dispersion relations. But still, it is
questionable how reliable the individual partial waves can be
separated.

The lack of data has led to the unpleasant situation that about half
of all nucleon and $\Delta$ resonances reported in the Karls\-ruhe
Helsinki analysis (KH80) \cite{Hohler:1979yr,Hohler:1993xq} and in
the Carnegie-Mellon analysis \cite{Cutkosky:1980rh} were not
confirmed in a later analysis at GWU \cite{Arndt:2006bf} using
larger and more precise data sets, in particular also spin rotation
parameters in the elastic pion-proton scattering from ITEP
\cite{Alekseev:1996gs,Alekseev:2000nk,Alekseev:2005zr}. Predictions
by \cite{Arndt:2006bf} of the backward asymmetry
\cite{Alekseev:2008cw} in the elastic pion-proton scattering favor
the GWU solution over the classical solutions of KH80 and CBM. So,
even in the case of $\pi N$ elastic scattering, the number of
contributing resonances is controversial, and concrete predictions
using the amplitudes of the KH80 and CBM analyses show significant
discrepancies when compared to data.

In photoproduction of hyperons, even more experiments are needed to
construct amplitudes from the data but experimental prospects are
better. As shown in \cite{Barker:1975bp,Chiang:1996em,Artru:2006xf},
precise measurements of eight carefully chosen observables are
sufficient to construct -- up to an overall phase -- the four
complex amplitudes describing photoproduction of a baryon with
spin-parity $J^P=1/2^+$ and a pseudoscalar meson. Measurements are
then needed of the differential cross section, experiments with
polarized photons yielding the beam asymmetry $\Sigma$, measurements
of the hyperon polarization $P$, of the target asymmetry $T$, and
double polarization experiments in which the polarization transfer
from the initial photon or of the target proton to the final state
hyperon is studied. This ambitious program is presently underway in
several laboratories but so far, data of a ``complete" experiment
are not yet available.

Even though the amplitudes cannot yet be constructed from existing
data, it is nevertheless possible to fit the data directly with
energy-dependent amplitudes. These amplitudes contain the physics:
the number and the properties of resonances. Several groups have
fitted existing data on hyperon production and reported which
resonances are required to achieve a good description of the data.

In a series of recent papers, we reported fits to data on pion and
photo-induced hyperon production: the exploratory study was
described in \cite{Anisovich:2010an,Anisovich:2011ye},
\cite{Anisovich:2011ka} was devoted to the study of a narrow
structure in $n\eta$, the results were summarized in
\cite{Anisovich:2011fc}, where also references to the data used can
be found. Interpretations of the results were given in
\cite{Anisovich:2011sv,Anisovich:2011su}. In this paper we compare
our amplitudes for hyperon and $\eta$ production with other PWA
models. Such a comparison allows for a deeper insight into the
model-dependence of the fits than a comparison of contributing
resonances and their properties can provide.

As mentioned above, the number of nucleon and $\Delta$ resonances in
the Karlsruhe-Helsinki analysis \cite{Hohler:1979yr,Hohler:1993xq}
and in the Carnegie-Mellon analysis \cite{Cutkosky:1980rh} is much
larger than in the analysis at GWU \cite{Arndt:2006bf}. In
\cite{Anisovich:2010an}, we have tested which impact this difference
has on the number of resonances observed in a multichannel analysis
which is constrained by either the Karlsruhe-Helsinki amplitudes or
by the GWU amplitudes. It was found that the number of required
resonances is the same and that the properties of resonances change
only little when the Karlsruhe-Helsinki amplitudes are replaced by
the GWU amplitudes. In order not to bias the analysis towards a
multitude of resonances, we use for our fits the GWU amplitudes

The paper is organized as follows: In the subsequent sections (2 and
3) we give a brief survey of data on hyperon and $\eta$ production
and of the fits which have been performed so far. In section 4 we
present the amplitudes for pion- and photo-produced production of
hyperons and of $\eta$ mesons. Our amplitudes for $\gamma p\to
p\pi^0$ and $n\pi^+$ have been presented elsewhere
\cite{Anisovich:2009zy}; however, the amplitude were based on the
solution BnGa2010. We show here the new amplitudes but refrain from
a renewed discussion since the new fits gave only small changes in
the amplitudes.

\section{Hyperon production}

Data on hyperon production by pion beams were taken in the 70ies and
80ies, in those glorious times when pion beams were still available
for experiments at all major laboratories for particle physics. Two
reactions are of particular importance:

i) production of $\Lambda$ hyperons in $\pi^-p\to \Lambda K^0$ gives
access to nucleon resonances. The reaction was studied from the
threshold region \cite{Baker:1977tg} up to 2.32\,GeV in invariant
mass \cite{Baker:1978qm}. The resonance parameters obtained in a
partial wave analysis \cite{Saxon:1979xu} of the data were found to
be in good agreement with those of $\pi N$ analyses. In the final
fits, masses and width of nucleon resonances were therefore fixed to
values derived from elastic $\pi N$ scattering; the main result was
thus the determination of $\pi N\to \Lambda K$ matrix elements.
Later measurements of the spin-rotation parameter \cite{Bell:1983dm}
confirmed the main predictions of~\cite{Saxon:1979xu}.

ii) Only isospin $I=3/2$ resonances can contribute to $\pi^+p\to
K^+\Sigma^+$. The reaction was studied by Candlin et al. from
threshold to 2.35\,GeV centre-of-mass energy \cite{Candlin:1982yv}.
An energy-depen\-dent partial wave analysis found a unique solution
giving a satisfactory fit to the data. Most resonances found in the
analyses of $\pi N$ elastic scattering were confirmed except some
one-star states \cite{Candlin:1983cw}. Spin rotation parameters were
measured later at two incident pion momenta \cite{Candlin:1988pn}.
The results were at most in fair agreement with predictions from
their earlier PWA \cite{Candlin:1983cw}. Hence their published
amplitudes are certainly not fully correct.

iii) The third reaction, $\pi^-p\to K^0\Sigma^0$, receives
contributions from both isospins \cite{Hart:1979jx}.

All three reactions benefit from the weak hyperon decay which
reveals the polarization status of the final-state baryon, i.e. the
recoil polarization.

Photo (and electro-)production of hyperons gives access to
additional quantities like helicity amplitudes and, in
electro-production, the dependence of transition amplitudes on the
(squared) momentum transfer. The latter variables are particulary
sensitive to the internal structure of baryon resonances. Again,
three reactions are relevant in this context: $\gamma p\to \Lambda
K^+$ and $\gamma p\to \Sigma^0 K^+$,
\cite{McNabb:2003nf,Zegers:2003ux,Bradford:2005pt,Bradford:2006ba,%
Lleres:2007tx,Lleres:2008em,McCracken:2009ra}, $\gamma p\to \Sigma^+
K^0$ \cite{Lawall:2005np,Hleiqawi:2007ad,Castelijns:2007qt}, and the
corresponding reactions off neutrons $\gamma n\to \Lambda K^0$,
$\gamma n\to \Sigma^0 K^0$, $\gamma n\to \Sigma^{\pm} K^{\mp}$. For
electroproduction, the $\gamma$ is replaced by a virtual photon
$\gamma^*$; we quote recent results here even though they are not
the topic of this paper \cite{Carman:2002se,Carman:2009fi}.

The conventional strategy to study the reaction dynamics and to
determine the properties of contributing resonances proceeds in
steps. In a first step, the $\pi N$ elastic scattering amplitudes
are determined in an energy-independent reconstruction. So far, this
step was carried out by three groups only \cite{Hohler:1979yr,%
Cutkosky:1980rh,Arndt:2006bf}. In a second step, a fit to the
resulting $\pi N$ elastic partial wave amplitudes is performed and
masses, widths and $\pi N$ couplings of contributing resonances are
determined. Reactions like $\pi^-p\to \Lambda K^0$ are used to
derive $\pi N\to \Lambda K^0$ transition matrix elements. In a third
step, helicity amplitudes $A_{1/2}$ and $A_{3/2}$ are specified by
including photoproduction data. Their $Q^2$ dependence, including
polarization transfer coefficients, follows from fits to
electroproduction data \cite{Aznauryan:2011qj,Tiator:2011pw}. In the
latter fits, hadronic properties of resonances are mostly frozen.

The Gie{\ss}en group was the first one which embarked the enterprize
to analyze simultaneously many reactions in a coupled channel fit.
The Giessen group starts from a chiral Lagrangian from which
background terms can be constructed with a minimum number of
parameters. The strong interaction parameters were frozen from
coupled-channel fits to pion-induced reactions and then,
photon-induced reactions were
included~\cite{Penner:2002ma,Penner:2002md}. The spin of resonances
was first limited to $3/2$; in more recent
studies~\cite{Shklyar:2004dy,Shklyar:2009cx}, spin $5/2$ resonances
have been included. The Gie\ss en group fits a large number of
different reactions like $\pi N\to\pi N,\,KY,$ $\eta N,\,\omega N$
and the corresponding photo-induced reactions. The recent
measurements high precision results from photo-production
experiments which include measurements of double-polarization
variables like $C_x,$ $C_z$ and $O_x$, $O_z$ were not yet available
at the time when the study was performed.

In many cases, photoproduction data were fitted using mas\-ses and
widths of known resonances as input. In this way, the effect of
coupled channels \cite{Chiang:2001pw} or of resonances in the
$u$-chan\-nel can be studied \cite{Janssen:2001pe} as well as the
impact of specific resonances
\cite{Chiang:2001as,Shklyar:2005xg,Shklyar:2006xw}; helicity
amplitudes of known resonances can be determined
\cite{delaPuente:2008bw,He:2009zzi}, and new resonances (the
so-called ``missing resonances") can be searched for
\cite{He:2008ty,Shyam:2009za}.  In the high-energy region, inclusion
of meson trajectories in the $t$-channel allows for a description of
various photoproduction channels over an extended energy range
\cite{Corthals:2006nz,Sibirtsev:2007wk,Sibirtsev:2009bj,Sibirtsev:2010yj}.
Elastic $\pi N$ scattering and the reaction $\pi^+ p \to K^+
\Sigma^+$ were described simultaneously in a unitary
coupled-channels approach by D\"oring et al. \cite{Doring:2010ap}.
The Bonn-J\"ulich approach starts  with a chiral Lagrangians and
takes dispersive parts of intermediate states fully into account.

\ Amplitudes for strangeness production are presented by Candlin~et
al. \cite{Candlin:1983cw}, by the Bonn-J\"ulich group
\cite{Doring:2010ap}, and by KAON-MAID \cite{Bennhold:1999nd}. The
three amplitudes are compared in section~\ref{Amplitudes}.

\section{Production of $\eta$ mesons}

Rather few data exist on the reaction $\pi^- p\to n\eta$, and
conflicting results have been reported \cite{Deinet:1969cd,%
Richards:1970cy,Bulos:1970zk,Debenham:1975bj,Feltesse:1975nz,%
Brown:1979ii,Crouch:1980vw,Prakhov:2005qb}. A lengthier discussion
on the reliability of different data sets is summarized
\cite{Durand:2008es}. In the BnGa analysis, these data are used to
constrain magnitudes for $N^*\to N \eta$ couplings. Hence we decided
to use the data from \cite{Richards:1970cy,Prakhov:2005qb} as main
input, and those in \cite{Brown:1979ii,Deinet:1969cd} with low
weight to control our solutions. The inconsistencies between the
different data sets demonstrate that techniques used in these -
rather old - experiments were not really adequate to identify $\eta$
mesons and to measure their yield.

The corresponding reaction with photons, $\gamma p\to p\eta$, has
been studied in a number of high-statistics experiments; a survey of
early experiments can be found in \cite{Bartholomy:2007zz}. The data
are mostly consistent, with some deviations between the data from
CLAS at Jlab \cite{Dugger:2002ft,Williams:2009yj} and Crystal Barrel
at ELSA; we decided to use only high statistics data with direct
detection of the $\eta$ mesons \cite{Bartholomy:2007zz,McNicoll:2010qk,%
Crede:2009zzb,Ajaka:1998zi,Bartalini:2007fg,Elsner:2007hm}.
Electroproduction has been studied as well; here we give reference
to the latest publication only \cite{:2008ff} in which earlier
experiments are quoted.

In experiments on $\eta$ production, the polarization state of the
outgoing nucleon is difficult to measure (and so far, was never
determined). Hence the corresponding amplitudes suffer from
considerably larger uncertainties.
\begin{figure*}
\begin{center}
\begin{tabular}{cc}
\hspace{-2mm}\includegraphics[width=0.42\textwidth]{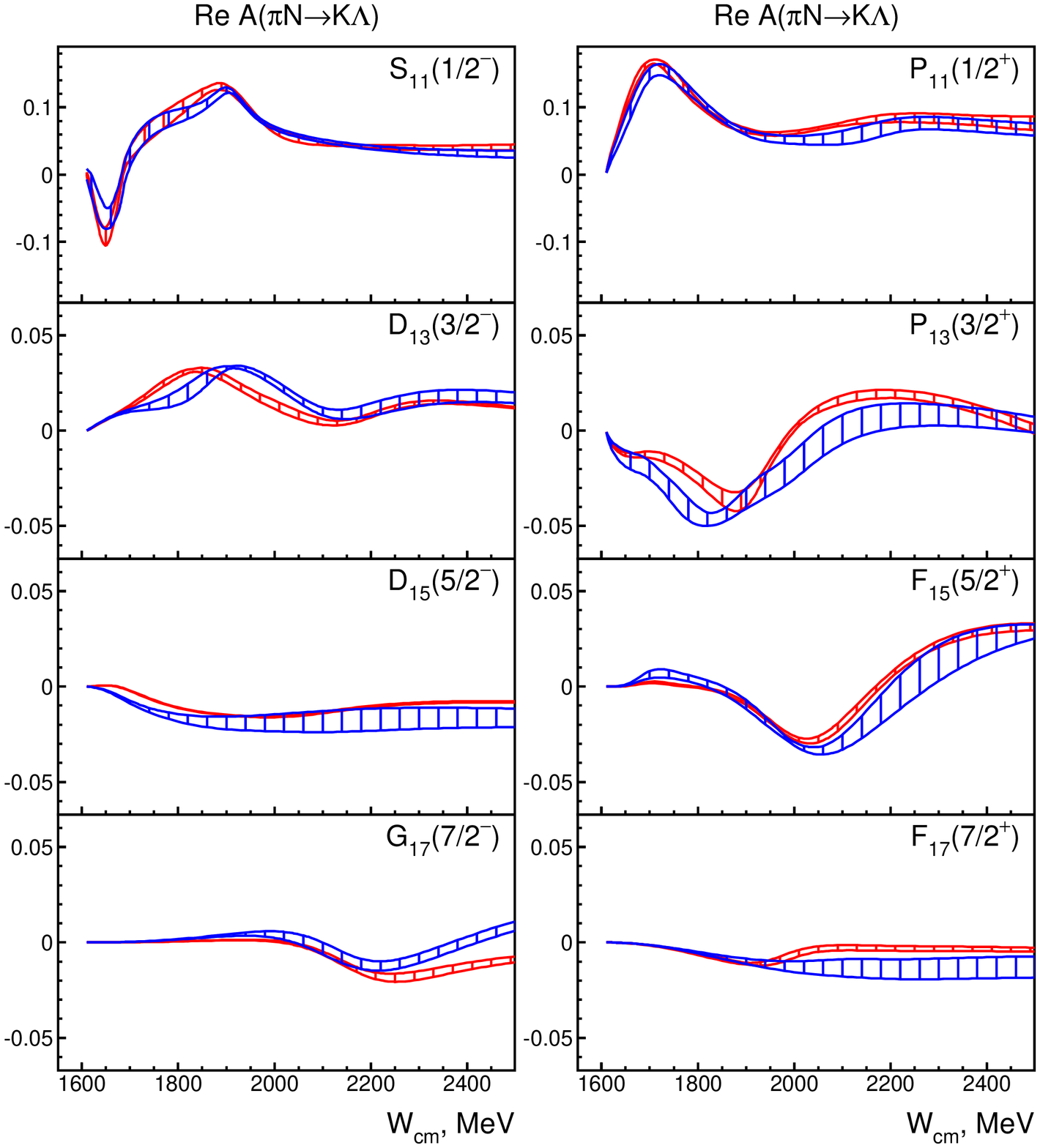}&
\hspace{-2mm}\includegraphics[width=0.42\textwidth]{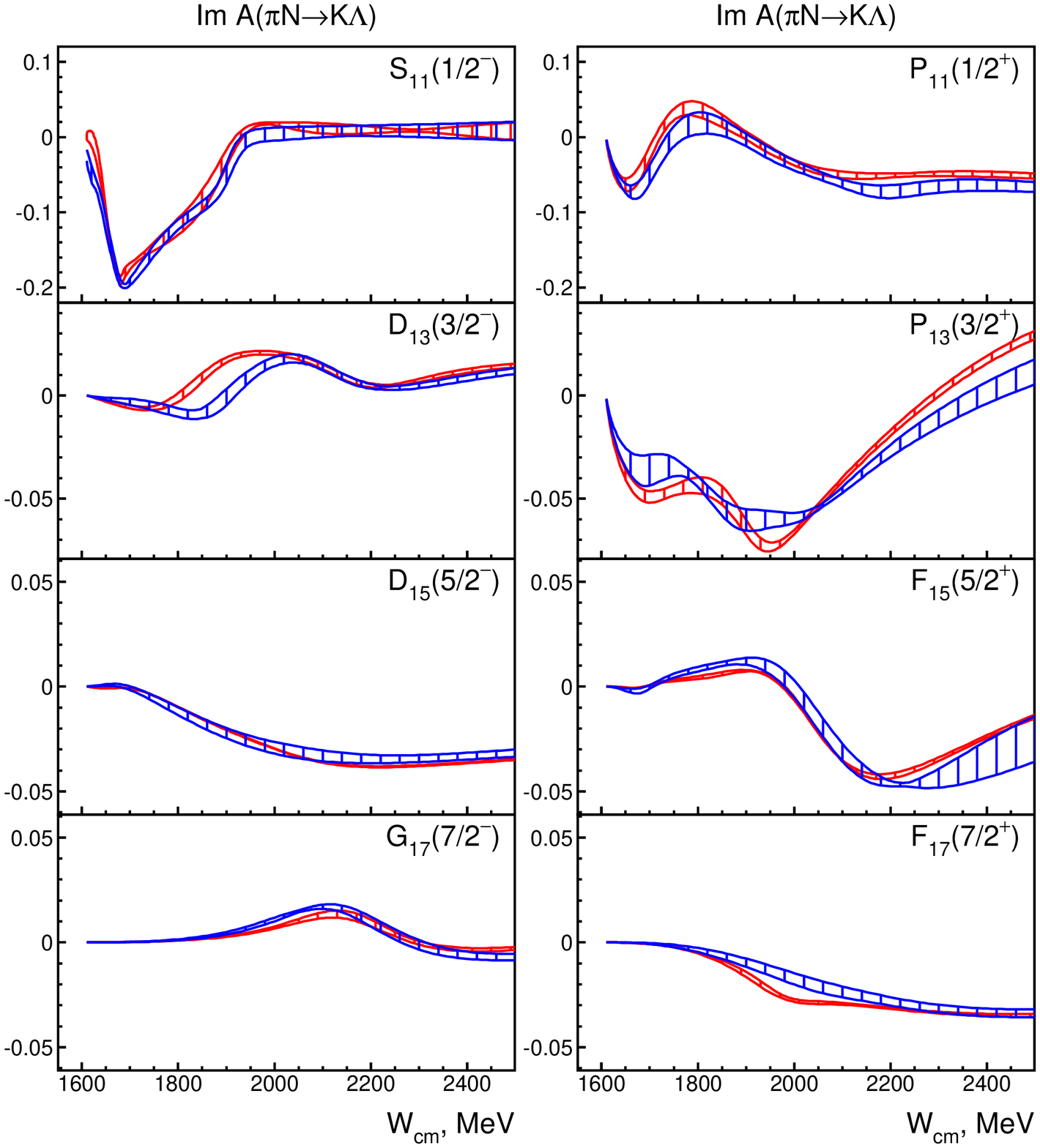}\\
\hspace{-2mm}\includegraphics[width=0.42\textwidth]{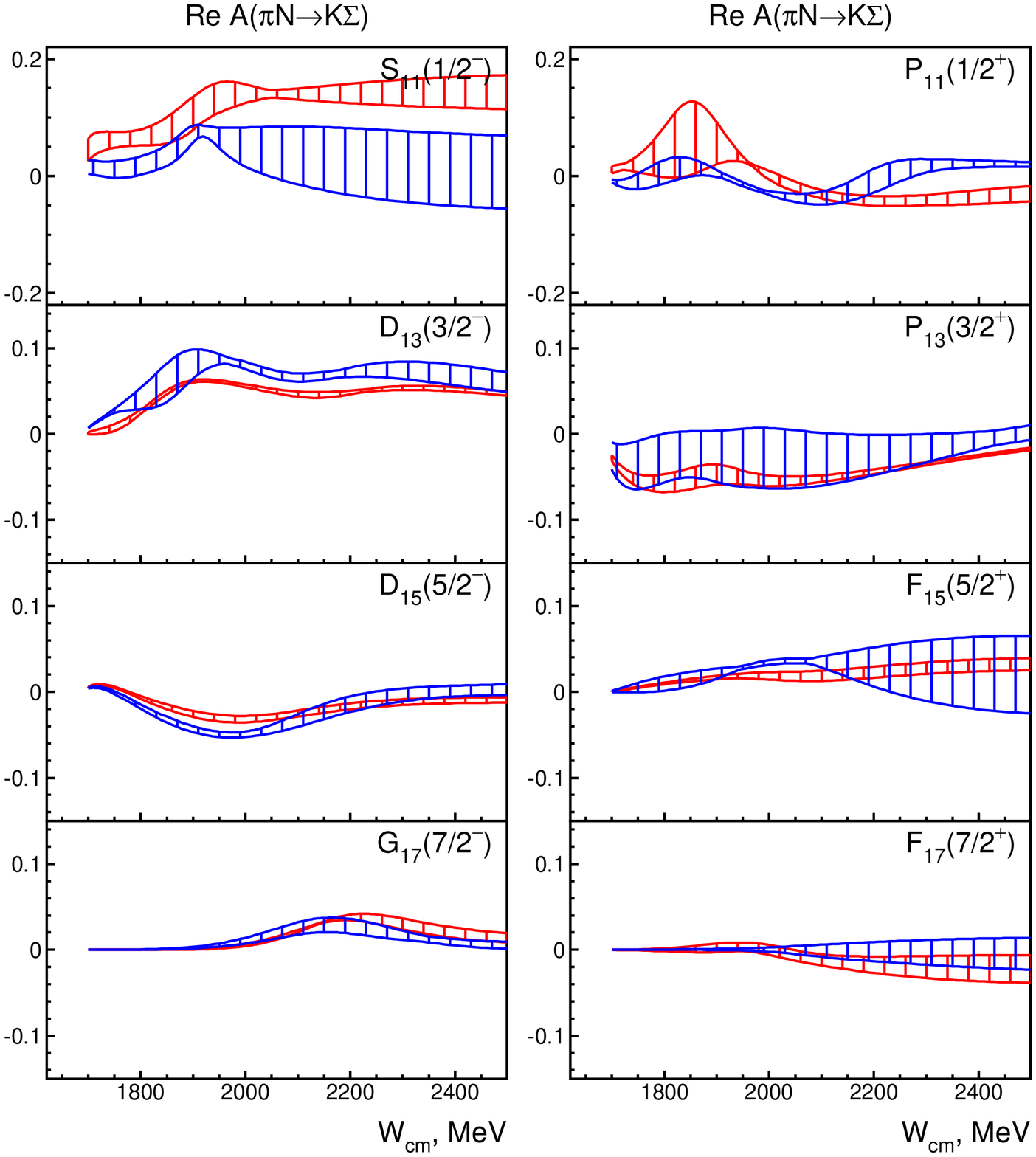}&
\hspace{-2mm}\includegraphics[width=0.42\textwidth]{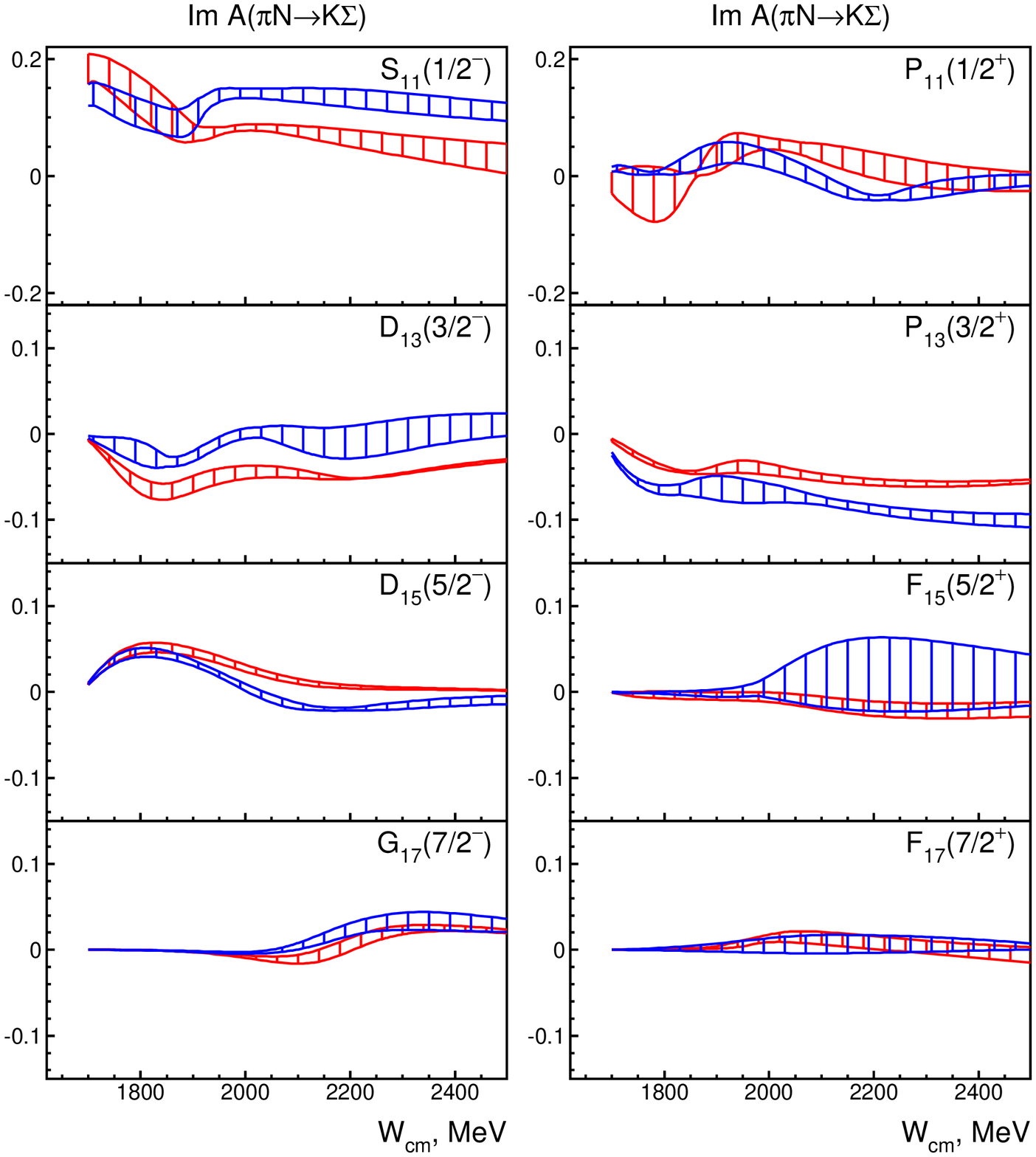}
\end{tabular}
\vspace{-2mm}\end{center} \caption{\label{piLambdaK}Decomposition of
the $\pi^- p\to \Lambda K^0$ (top) and $\pi N\to N\Sigma$
isospin-1/2 (bottom) transition amplitudes. The light (red) shaded
areas give the range from a variety of different fits derived from
solution BnGa2011-01, the dark (blue) shaded area from solution
BnGa2011-02.}\vspace{-2mm}
\end{figure*}

\begin{figure*}
\begin{center}
\begin{tabular}{cc}
\hspace{-2mm}\includegraphics[width=0.42\textwidth]{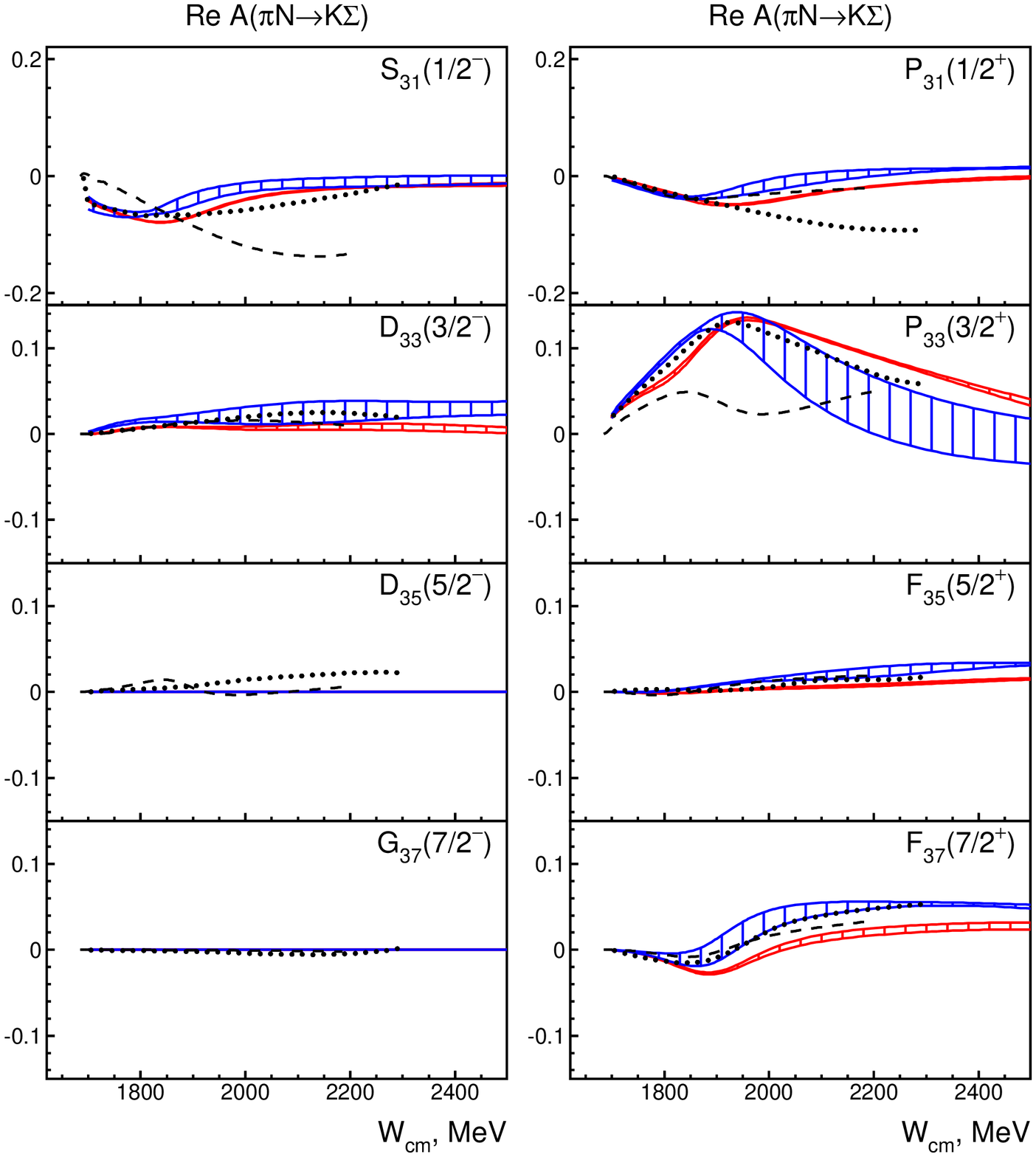}&
\hspace{-2mm}\includegraphics[width=0.42\textwidth]{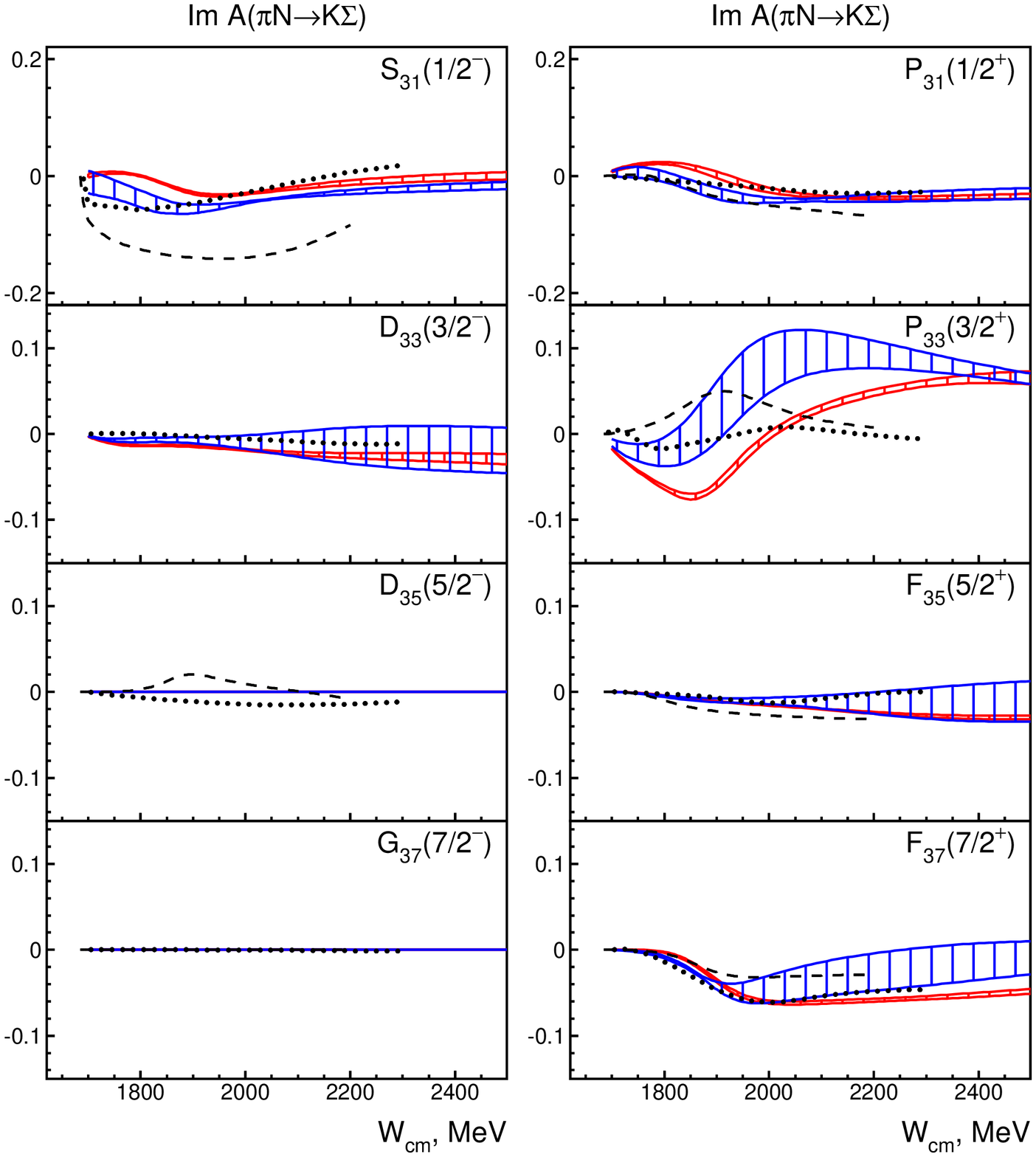}\\
\hspace{-2mm}\includegraphics[width=0.42\textwidth]{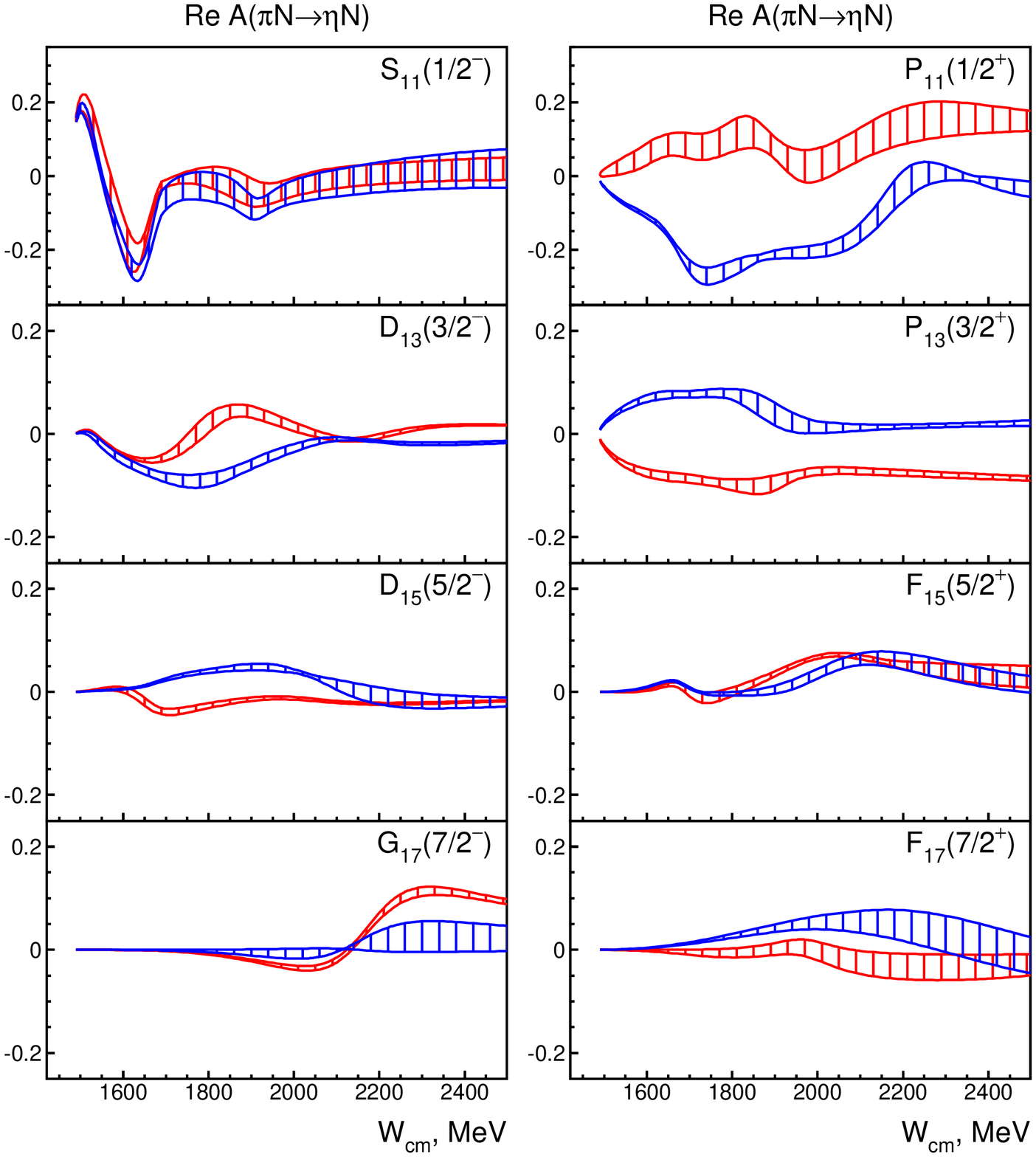}&
\hspace{-2mm}\includegraphics[width=0.42\textwidth]{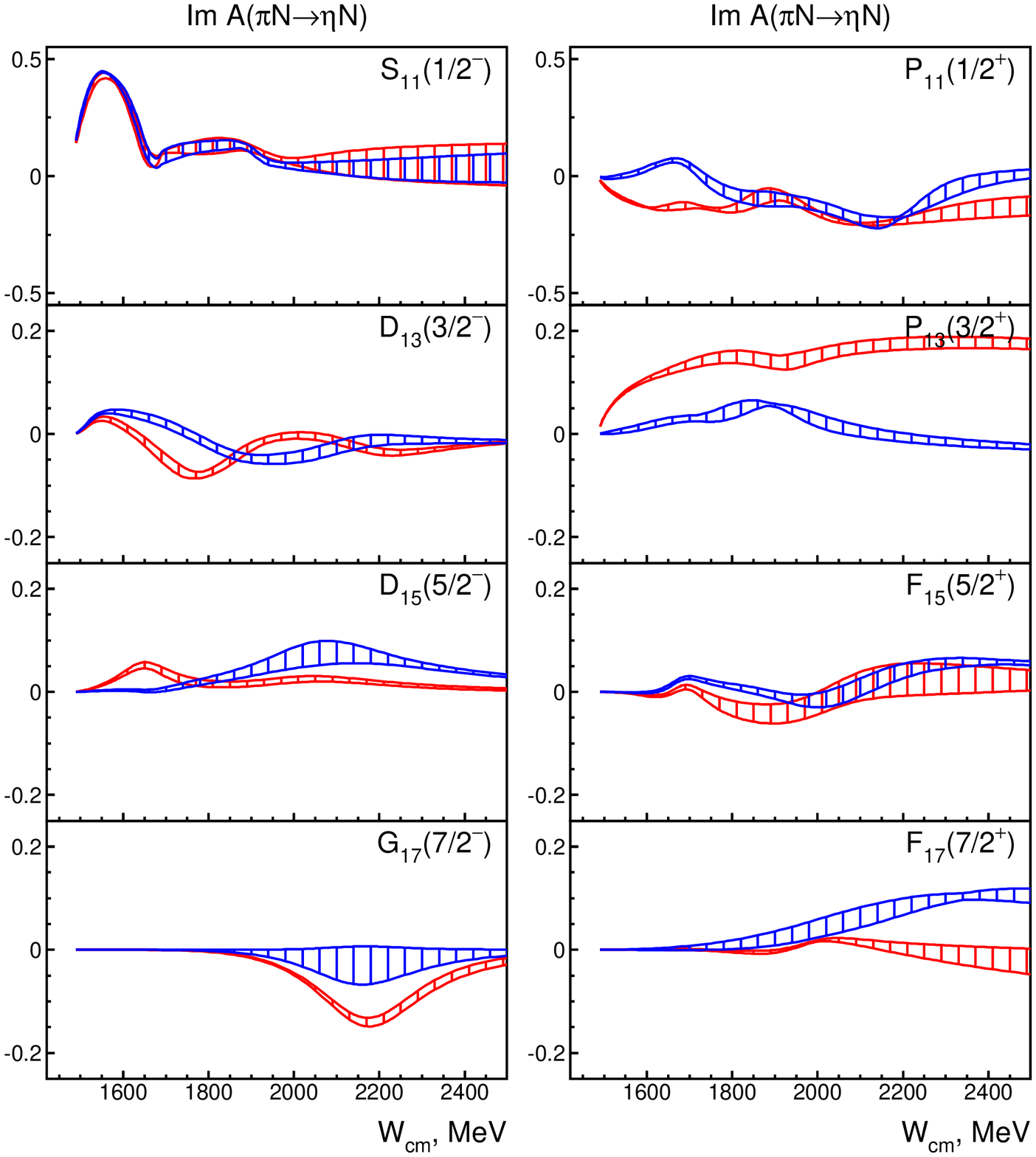}
\end{tabular}
\vspace{-2mm}\end{center} \caption{\label{pineta}Decomposition of
the $\pi N\to N\Sigma$ isospin-3/2 (top) and  $\pi^- p\to n\eta$
transition amplitudes. The light (red) shaded areas give the range
from a variety of different fits derived from solution BnGa2011-01,
the dark (blue) shaded area from solution BnGa2011-02. The dotted
curve represents the PWA from \cite{Candlin:1983cw}, dashed one from
\cite{Doring:2010ap}.}
\begin{minipage}[t]{0.60\textwidth}
\begin{center}
\begin{tabular}{cc}
\hspace{-2mm}\includegraphics[width=0.48\textwidth,height=0.45\textwidth]{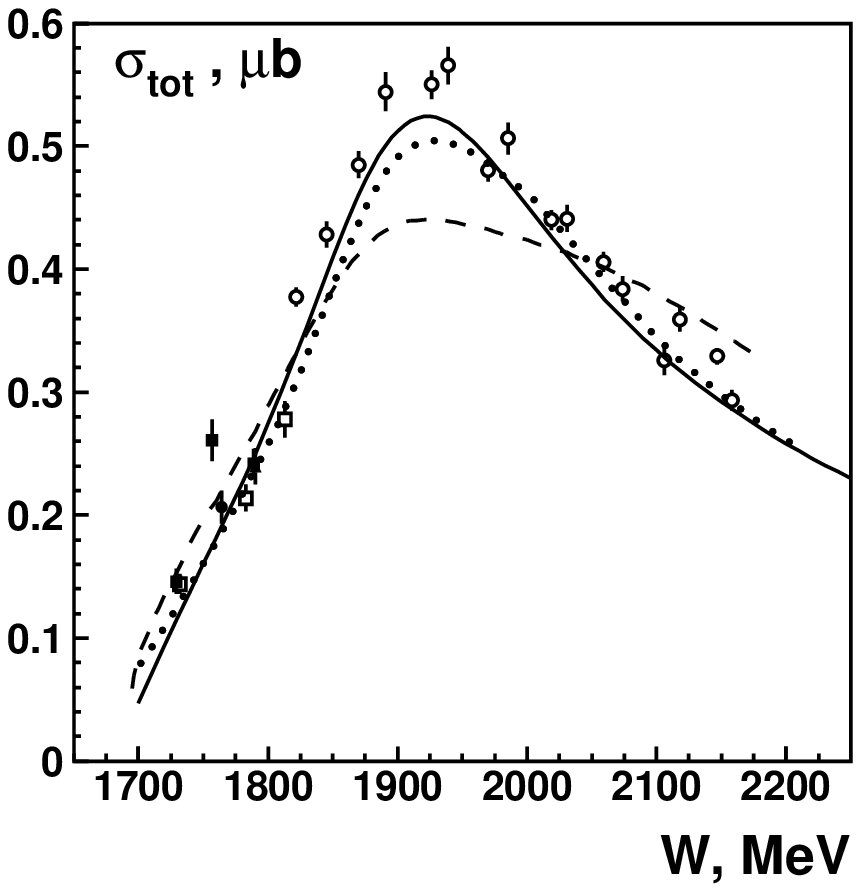}&
\hspace{-4mm}\includegraphics[width=0.48\textwidth,height=0.45\textwidth]{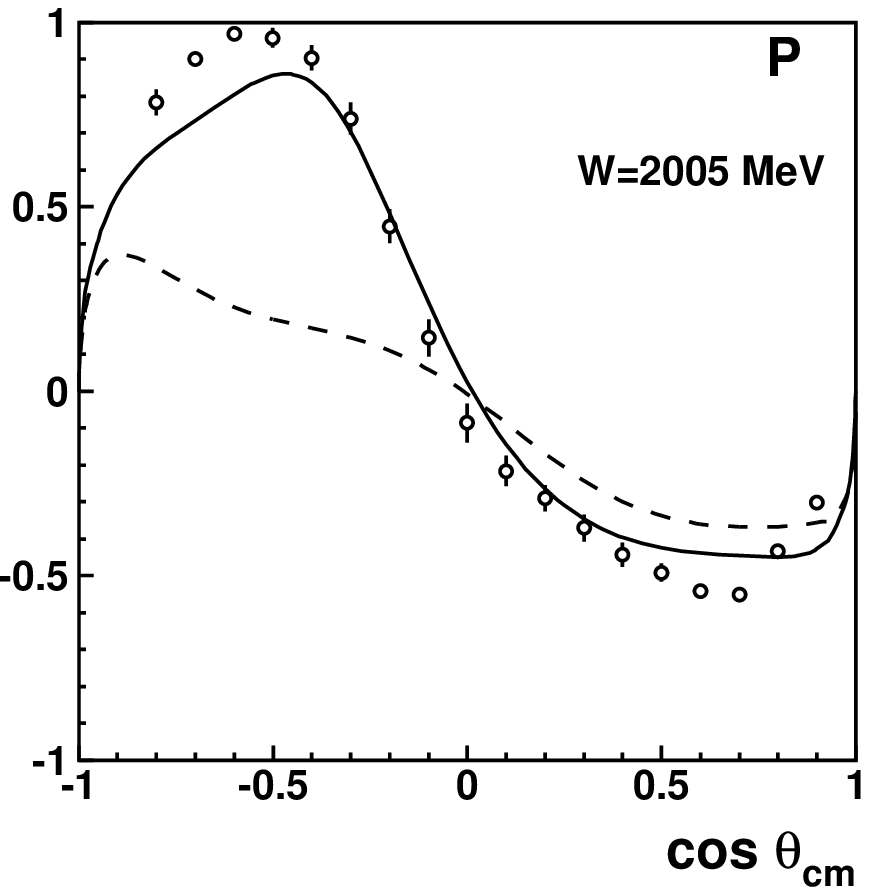}
\end{tabular}
\end{center}
\end{minipage}
\hspace{2mm}\begin{minipage}[t]{0.38\textwidth}
\caption{\label{piSig}Left: Cross section for the reaction $\pi^+
p\to \Sigma^+ K^+$;  Open circles \cite{Candlin:1982yv}, Black
squares \cite{Winik:1977mm}, Triangle - \cite{Baltay:1961}, Black
circles \cite{Crawford:1962zz}, Open squares
\cite{Carayannopoulos:1965}. (Some symbols overlap and are difficult
to recognize.) The solid curve represent this work, the dotted line
is from \cite{Candlin:1983cw}, and the dashed line from
\cite{Doring:2010ap}. Right: Recoil Polarization $P$ for $\gamma
p\to \Lambda K^+$ at 2\,GeV. Data from \cite{McCracken:2009ra} show
the statistical error. The solid curve is from the BnGa fit, the
dashed curve the prediction of KAON-MAID \cite{Bennhold:1999nd}.}
\end{minipage}
\end{figure*}
\section{\label{Amplitudes}Amplitudes}

The amplitudes presented here are derived from the BnGa multichannel
analysis of a large fraction of the world data on nucleon
resonances. Due to the complexity of the problem and missing
information on further polarization variables, the fits due not
converge to a well defined minimum. Instead, we found two classes of
solutions, called BnGa2011-01 and BnGa2011-02. These two classes of
solutions are defined by the number of resonances with $J^P=7/2^+$,
one has one $7/2^+$ resonance, the other one two. Within both
classes of solutions, a number of different solutions were found
which differ in smaller details. These provide error bands for the
amplitudes.

In Figs. \ref{piLambdaK} - \ref{pineta} we present our amplitudes
and compare them to earlier results. The amplitudes for $\pi^- p\to
\Lambda K^0$ in \cite{Saxon:1979xu} are given only in the form of
Argand diagrams without a mass scale, hence a comparison of the
amplitudes is not possible. The error bands of the two Bonn-Gatchina
solutions and their consistency is relatively good even though
considerable differences can be seen for smaller amplitudes. These
differences can be qualitative like for the $J^P=3/2^+$ wave where
the observed pattern seems to be shifted in energy; mass and width
of the resonances in this wave optimize, however, for nearly the
same pole positions; only the relative phases between consecutive
resonances are different. In contrast, the (small) $7/2^+$ wave
shows a significant structure in solution BnGa2011-01 which is
absent in BnGa2011-02.

For the $\pi N\to \Sigma K$ reaction, there is a much wider spread
of results. For isospin 1/2, for nucleon excitations, no comparison
with other work is available. The amplitudes for isospin 3/2 are
shown in Fig. \ref{pineta}, upper panel. The amplitudes are from the
work of Candlin {\it et al.} \cite{Candlin:1983cw} and from the
recent Bonn/J\"ulich analysis \cite{Doring:2010ap}. Candlin {\it et
al.} and our amplitudes are in fair agreement; those from
Bonn/J\"ulich show partly larger discrepancies. In particular, the
Bonn/J\"ulich $S_{31}(1/2^-)$ waves deviates significantly from the
ones reconstructed by Candlin {\it et al.} and from our amplitude,
even at the lowest energies. The $P_{33}(3/2^+)$ partial wave seems
to be not very well defined by the data, sizable differences are
seen between our two solutions, the Bonn/J\"ulich analysis, and the
solutions from the work of Candlin {\it et al.} where the imaginary
part is comparatively small. The $P_{31}(1/2^+)$ wave from Candlin
{\it et al.} disagrees with the findings by Bonn/J\"ulich and our
findings: in this wave, no resonance is supposed to contribute in
the work of Candlin {\it et al.} while in the other two analyses,
one $P_{31}(1/2^+)$ resonances is used to describe the data. This
work finds no contribution from the $D_{35}(5/2^-)$ wave while
Candlin {\it et al.} and Bonn/J\"ulich identify a small
contribution. In the higher partial waves, no structural differences
are seen.

To trace the differences we have reconstructed the total cross
section from the sets of three partial wave amplitudes. These are
compared with the data in Fig. \ref{piSig}. It is obvious, that some
intensity is missed in the amplitudes of \cite{Doring:2010ap}.
Dynamical coupled-channels models based on effective chiral
Lagrangians have the advantage of a much reduced number of fit
parameters since they provide a microscopical description of the
background \cite{Doring:2009bi,Doring:2009yv}. In some cases,
resonances can even be constructed from the iteration of background
terms \cite{Kaiser:1995cy,Meissner:1999vr,Nieves:2001wt}. However,
there is the possibility that not all background amplitudes provided
by Nature are included in the calculation. It is then unclear to
which minimum the fit converges.

The reaction $\pi N\to \eta N$ lacks good data; polarization data
are not available. In some partial waves, even the sign of coupling
constants is not defined. Nevertheless, the data are useful since
they constrain the $N\eta$ decay modes of nucleon resonances. The
signs of the two partial wave amplitudes $P_{11}(1/2^+)$ and
$P_{13}(3/2^+)$ can both flip; this ambiguity leads to ambiguities
in the $N\eta$ contribution of $N(1710)1/2^+$ and $N(1720)3/2^+$.
Measurements of the helicity dependence of the $\eta$
photoproduction cross section should thus help to resolve this
ambiguity.

\begin{figure*}
\begin{center}
\begin{tabular}{cc}
\hspace{-2mm}\includegraphics[width=0.42\textwidth]{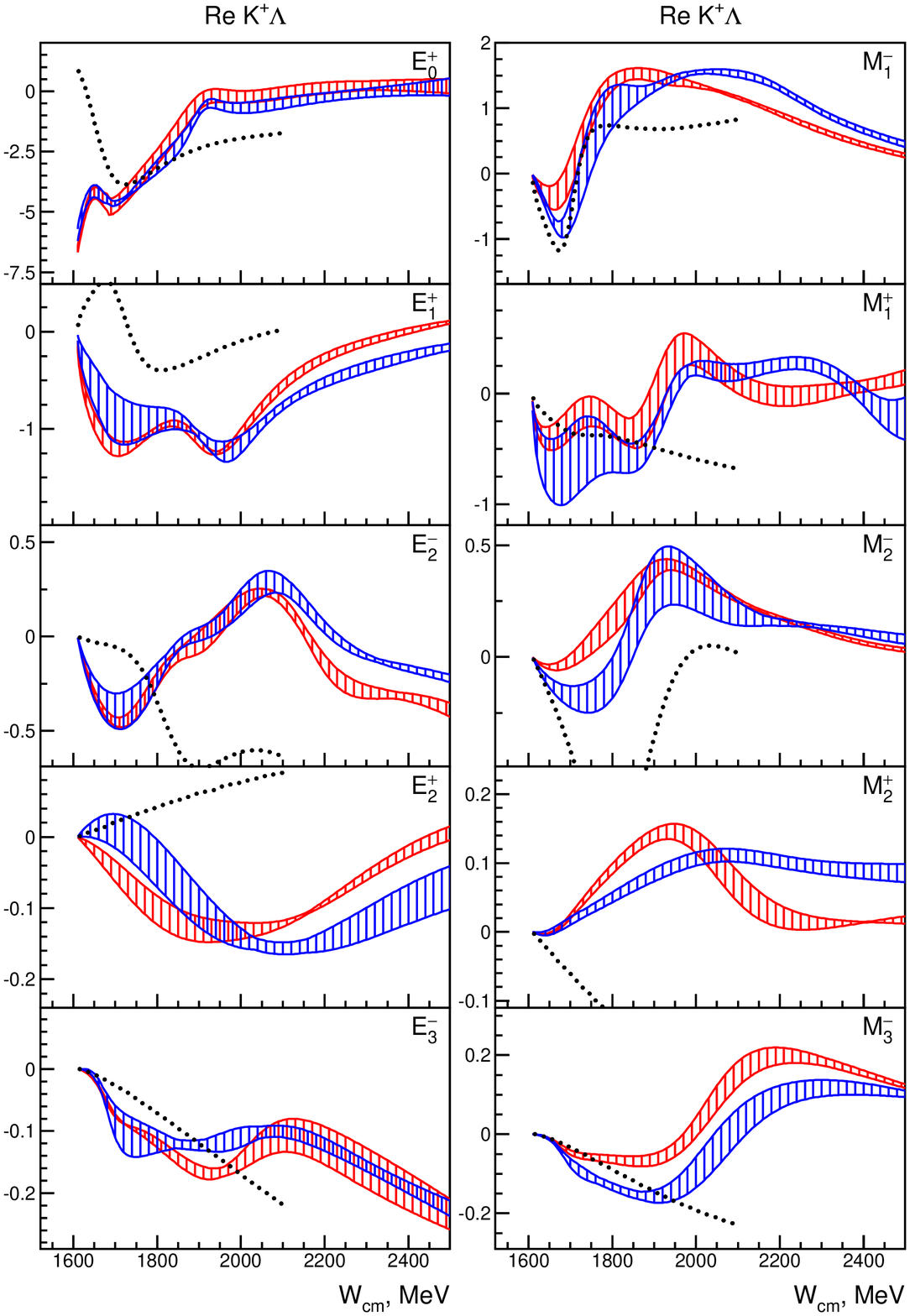}&
\hspace{-2mm}\includegraphics[width=0.42\textwidth]{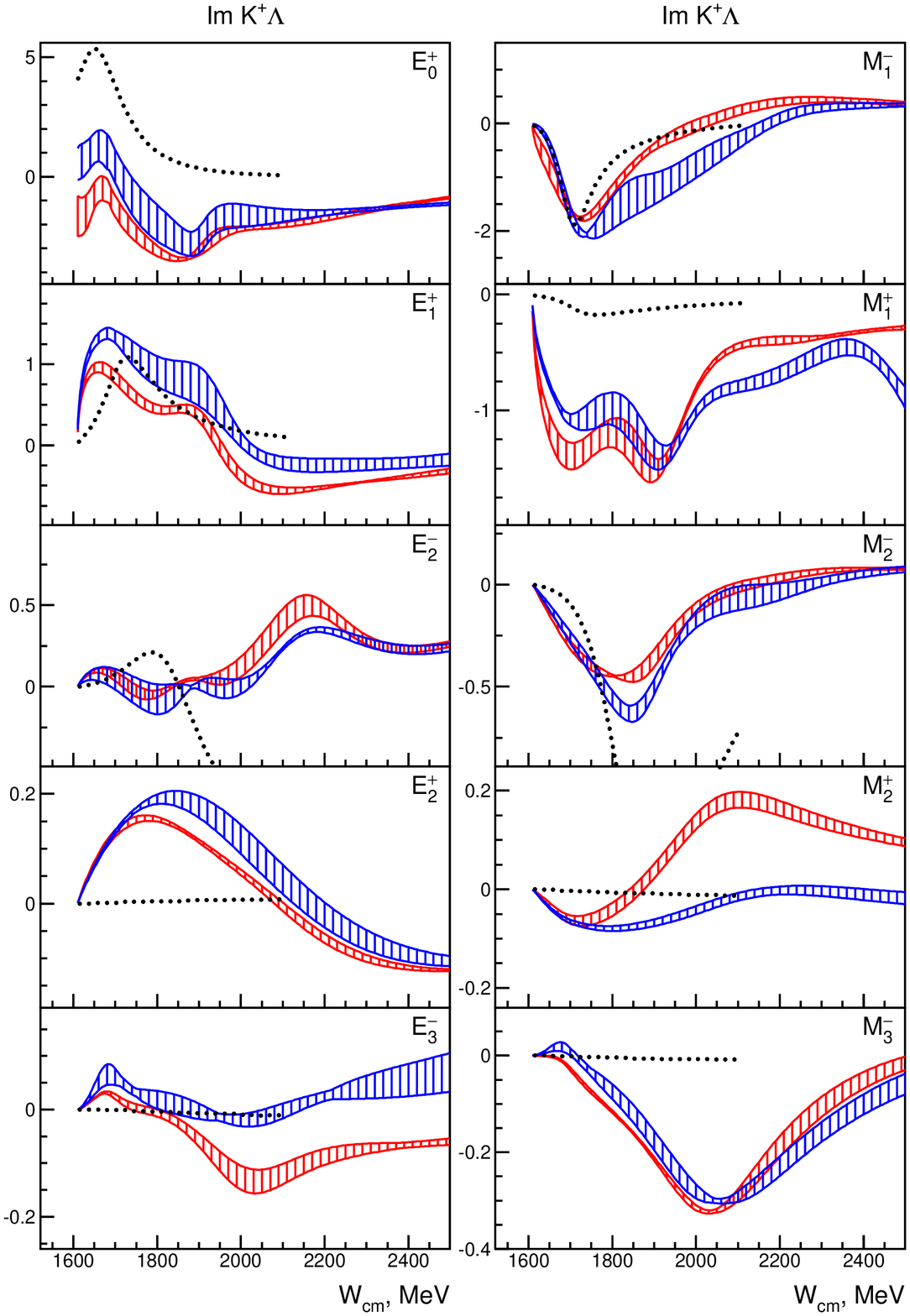}\\
\hspace{-2mm}\includegraphics[width=0.42\textwidth]{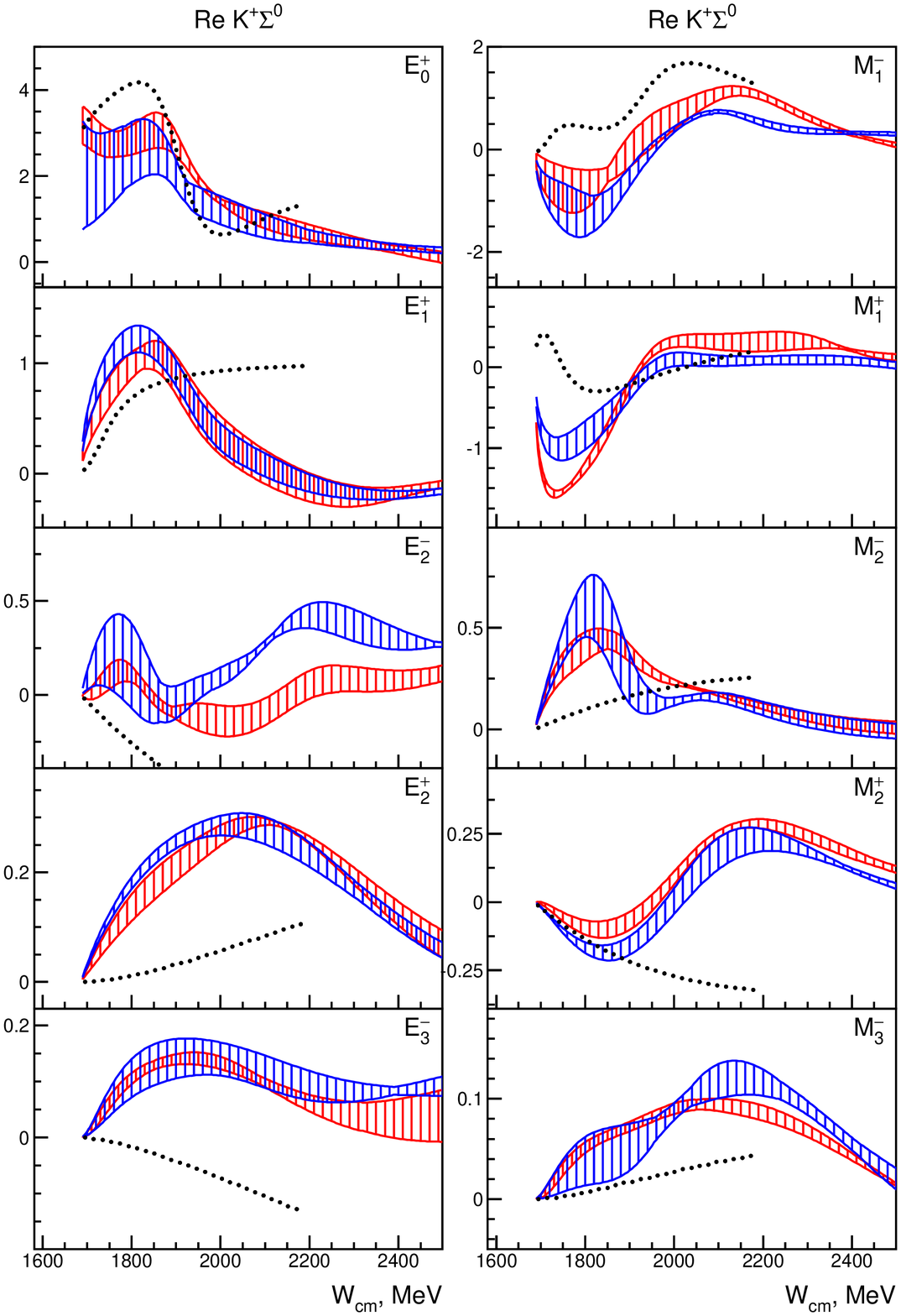}&
\hspace{-2mm}\includegraphics[width=0.42\textwidth]{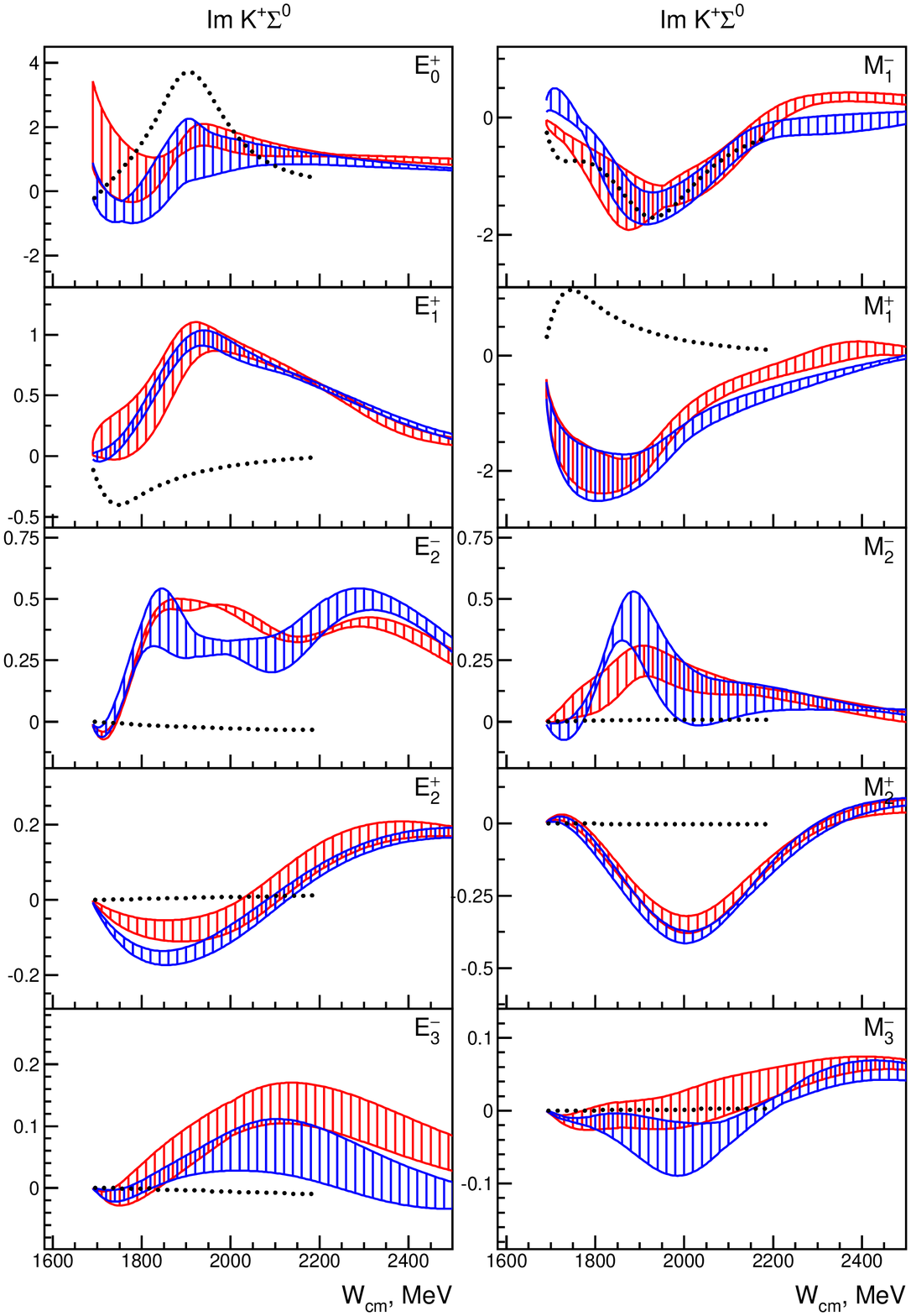}
\end{tabular}
\end{center}
\caption{\label{gLambdaK}Multipole decomposition of the $\gamma p\to
\Lambda K^+$ (top) and $\gamma p\to \Sigma^0 K^+$ (bottom)
transition amplitudes. The multipoles are given in units of
$10^{-3}/M_{\pi^+}$. The light (red) shaded areas give the range
from a variety of different fits derived from solution BnGa2011-01,
the dark (blue) shaded area from solution BnGa2011-02. The dotted
line represents the fit from KAON-MAID \cite{Bennhold:1999nd}.}
\end{figure*}
\begin{figure*}
\begin{center}
\begin{tabular}{cc}
\hspace{-2mm}\includegraphics[width=0.42\textwidth]{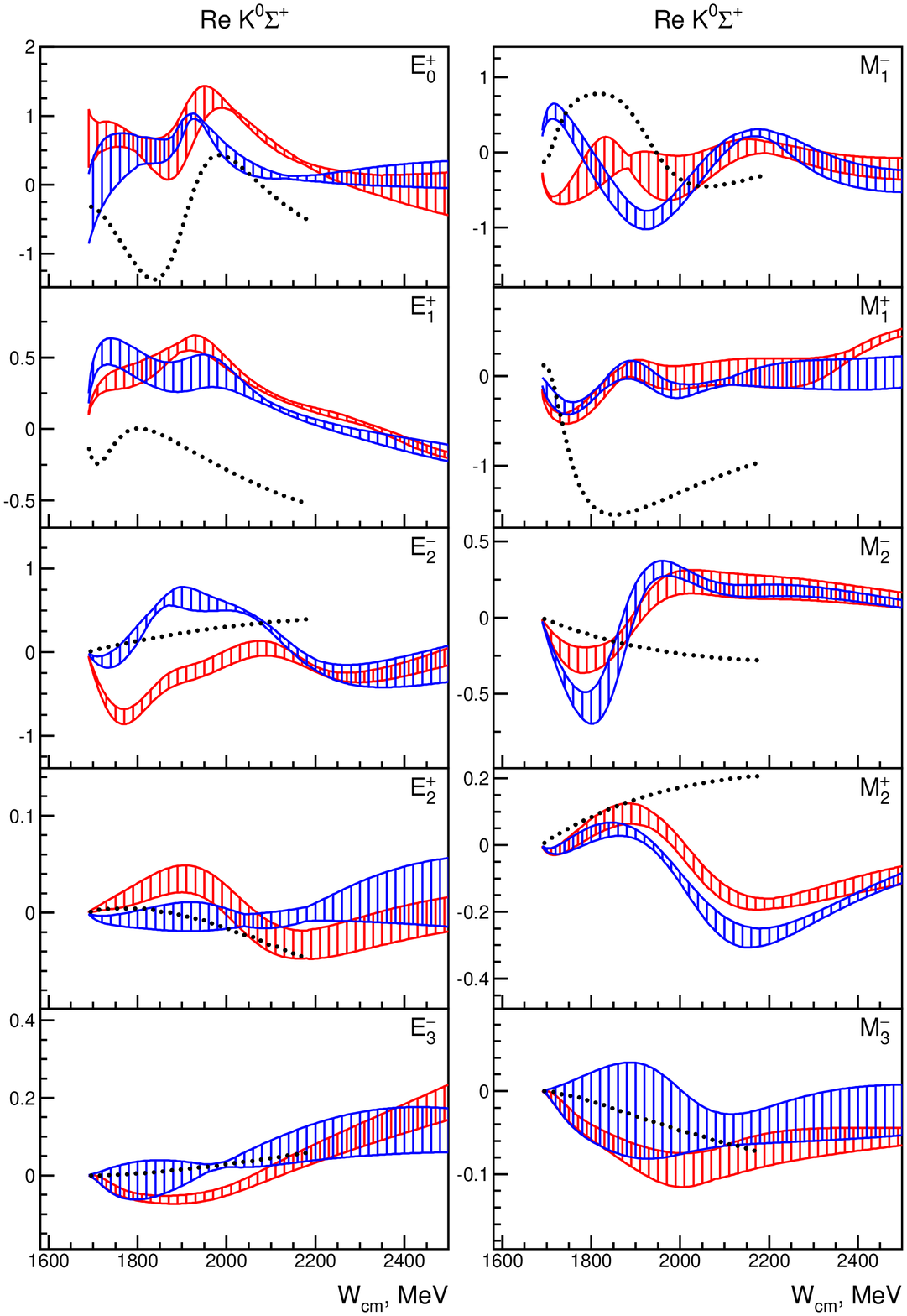}&
\hspace{-2mm}\includegraphics[width=0.42\textwidth]{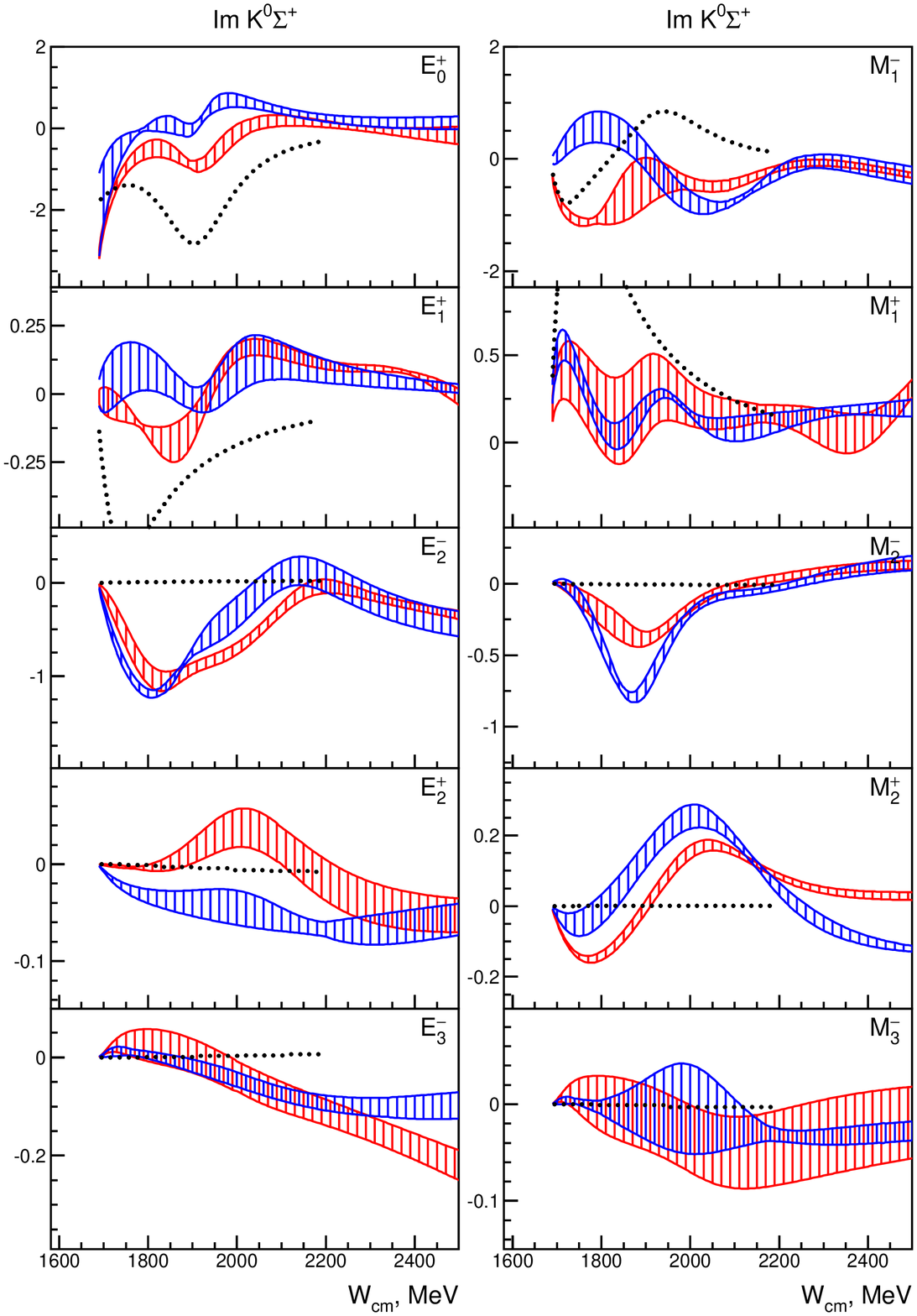}\\
\hspace{-2mm}\includegraphics[width=0.42\textwidth]{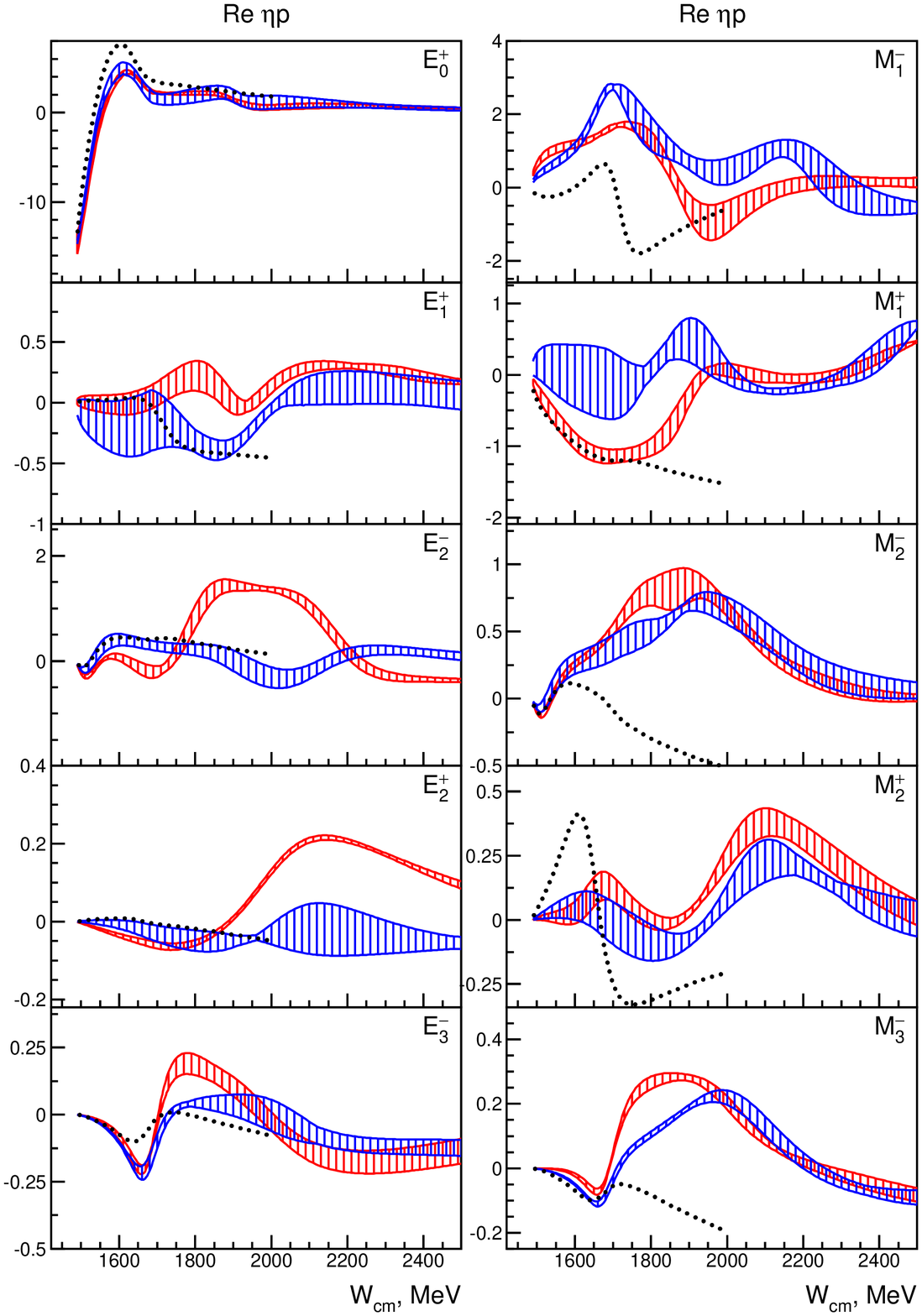}&
\hspace{-2mm}\includegraphics[width=0.42\textwidth]{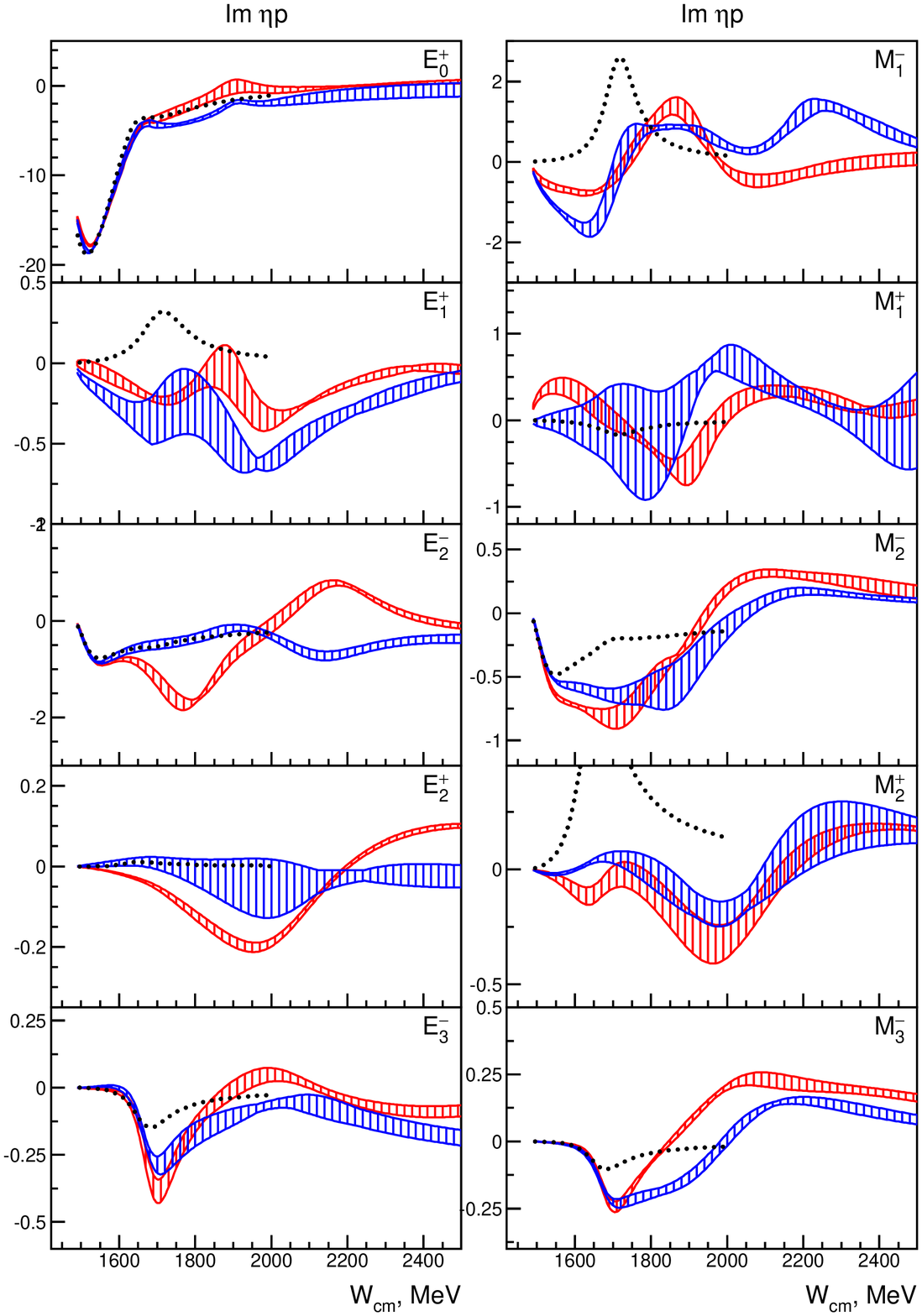}
\end{tabular}
\end{center}
\caption{\label{gppeta}Multipole decomposition of the $\gamma p\to
\Sigma^+ K^0$ (top) and $\gamma p\to p\eta$ (bottom) transition
amplitudes. The multipoles are given in units of
$10^{-3}/M_{\pi^+}$. The light (red) shaded areas give the range
from a variety of different fits derived from solution BnGa2011-01,
the dark (blue) shaded area from solution BnGa2011-02. The dotted
line represents the fit from KAON-MAID.}
\end{figure*}
\begin{figure*}
\begin{center}
\begin{tabular}{cc}
\hspace{-2mm}\includegraphics[width=0.42\textwidth]{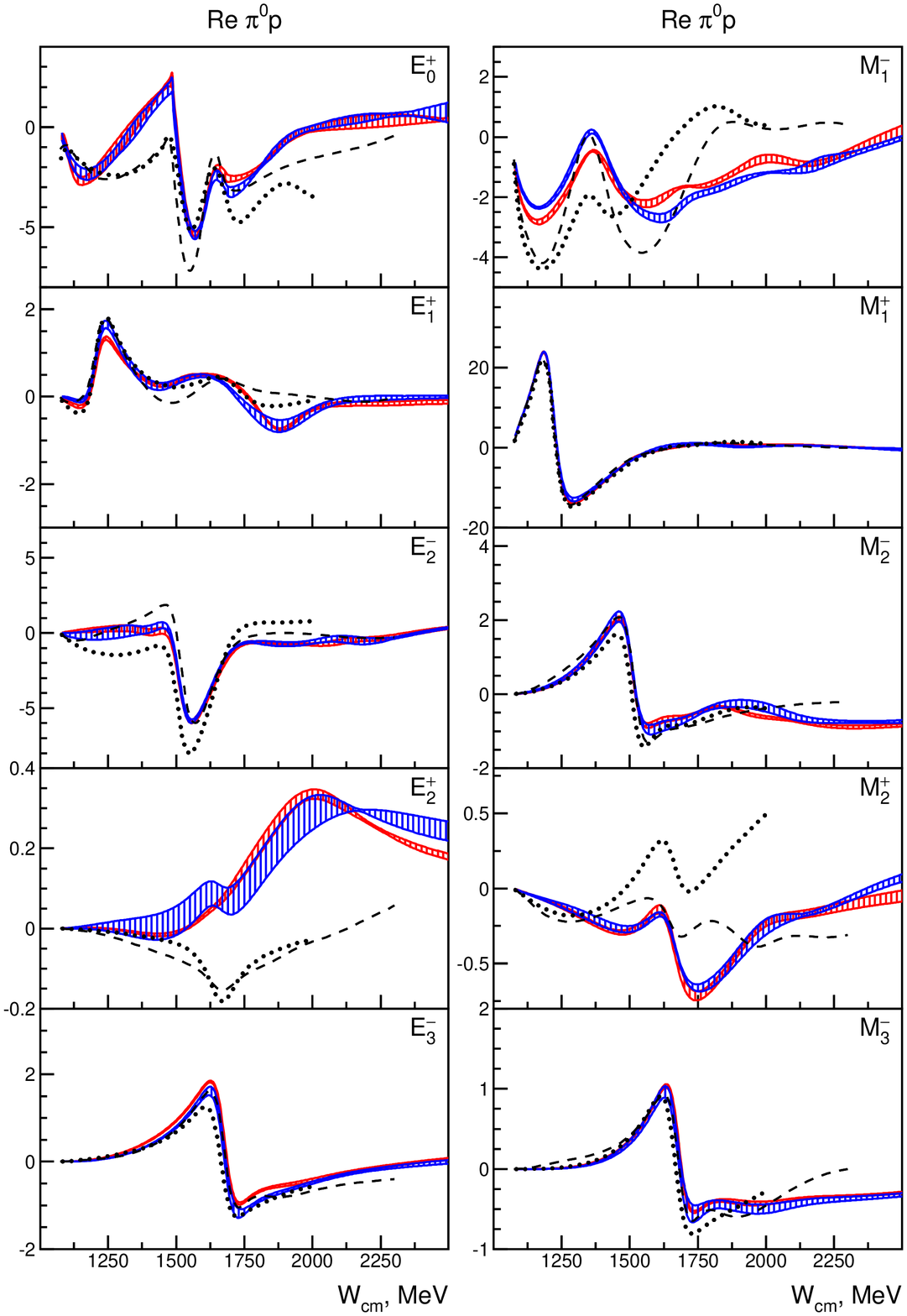}&
\hspace{-2mm}\includegraphics[width=0.42\textwidth]{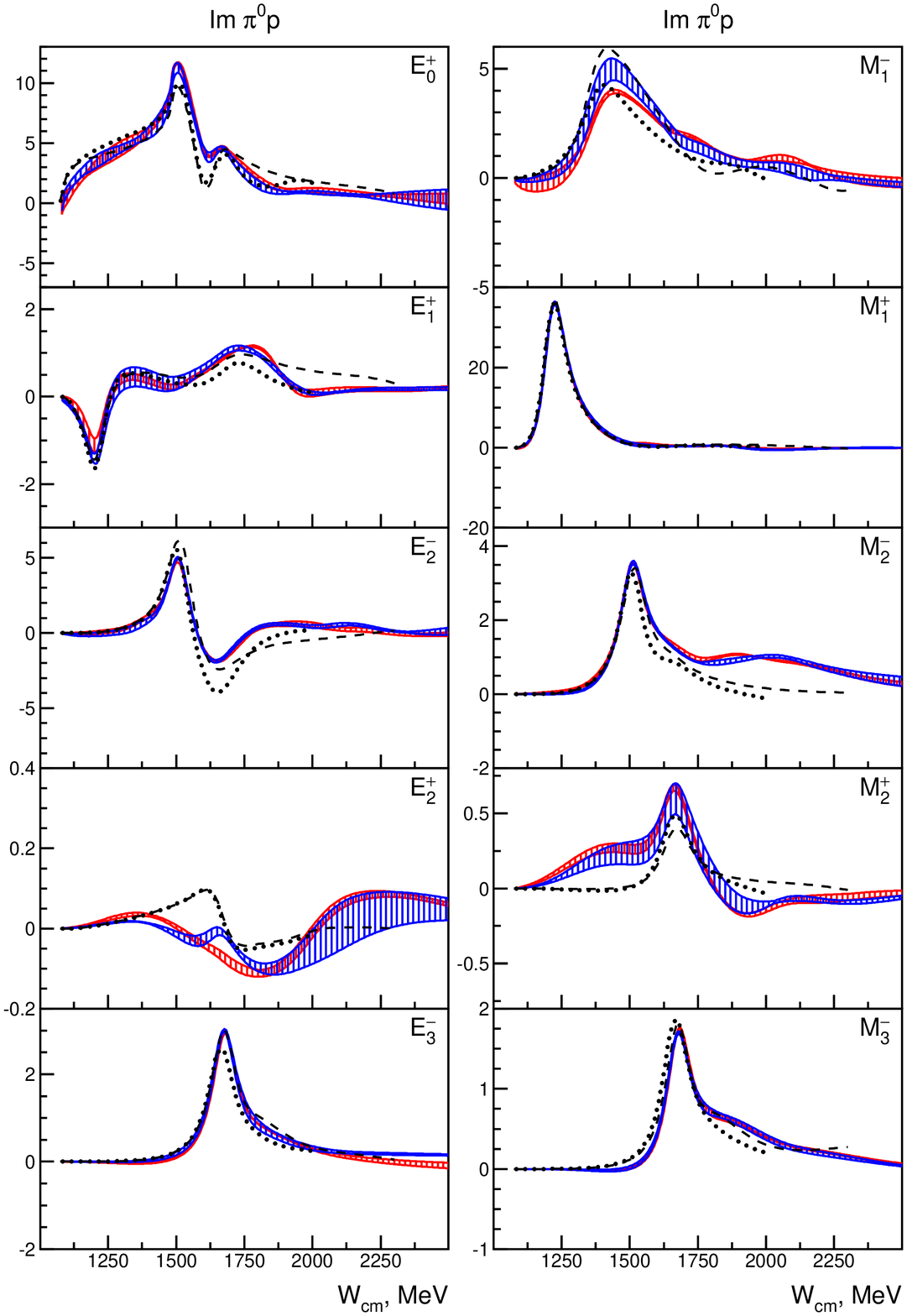}\\
\hspace{-2mm}\includegraphics[width=0.42\textwidth]{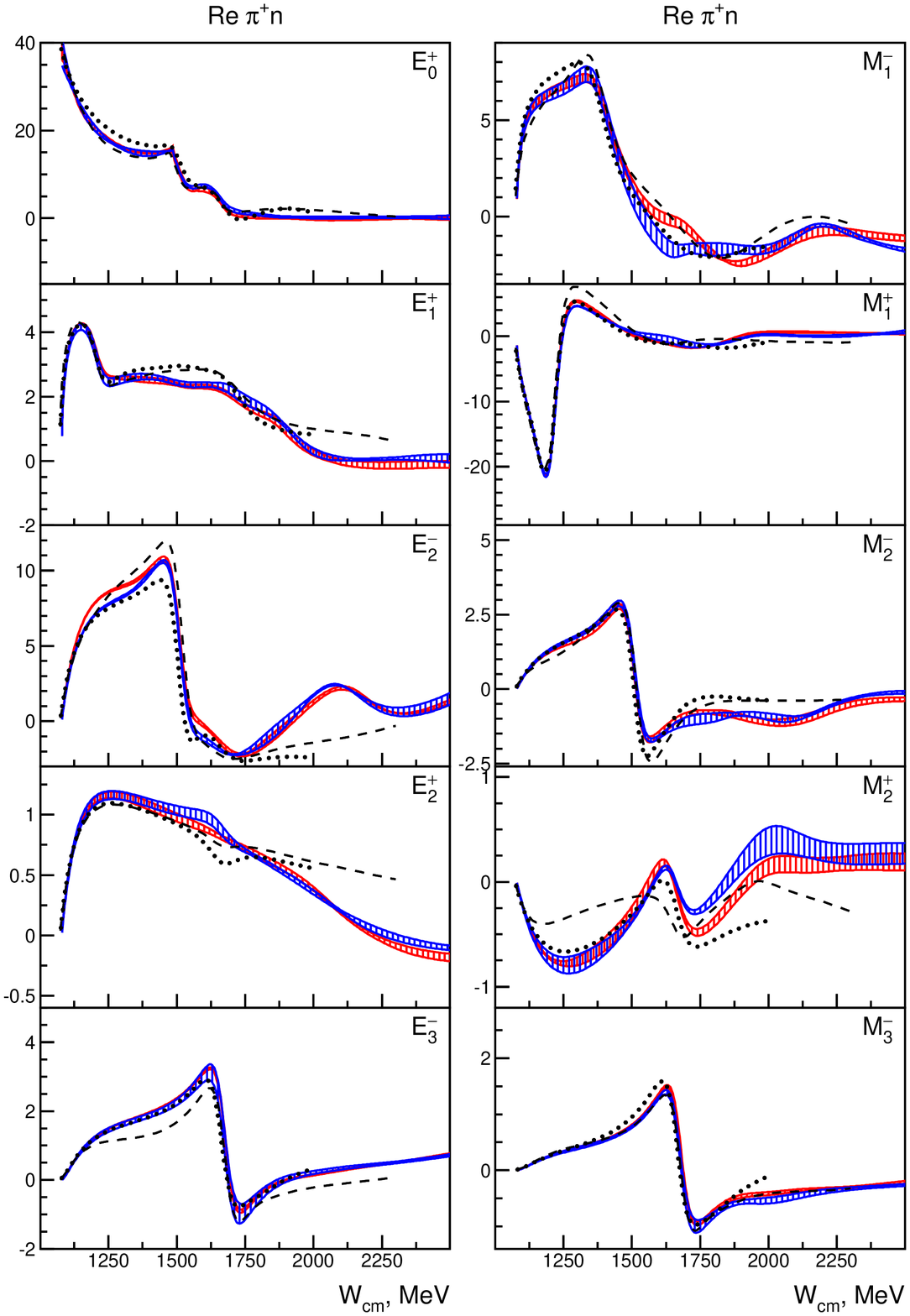}&
\hspace{-2mm}\includegraphics[width=0.42\textwidth]{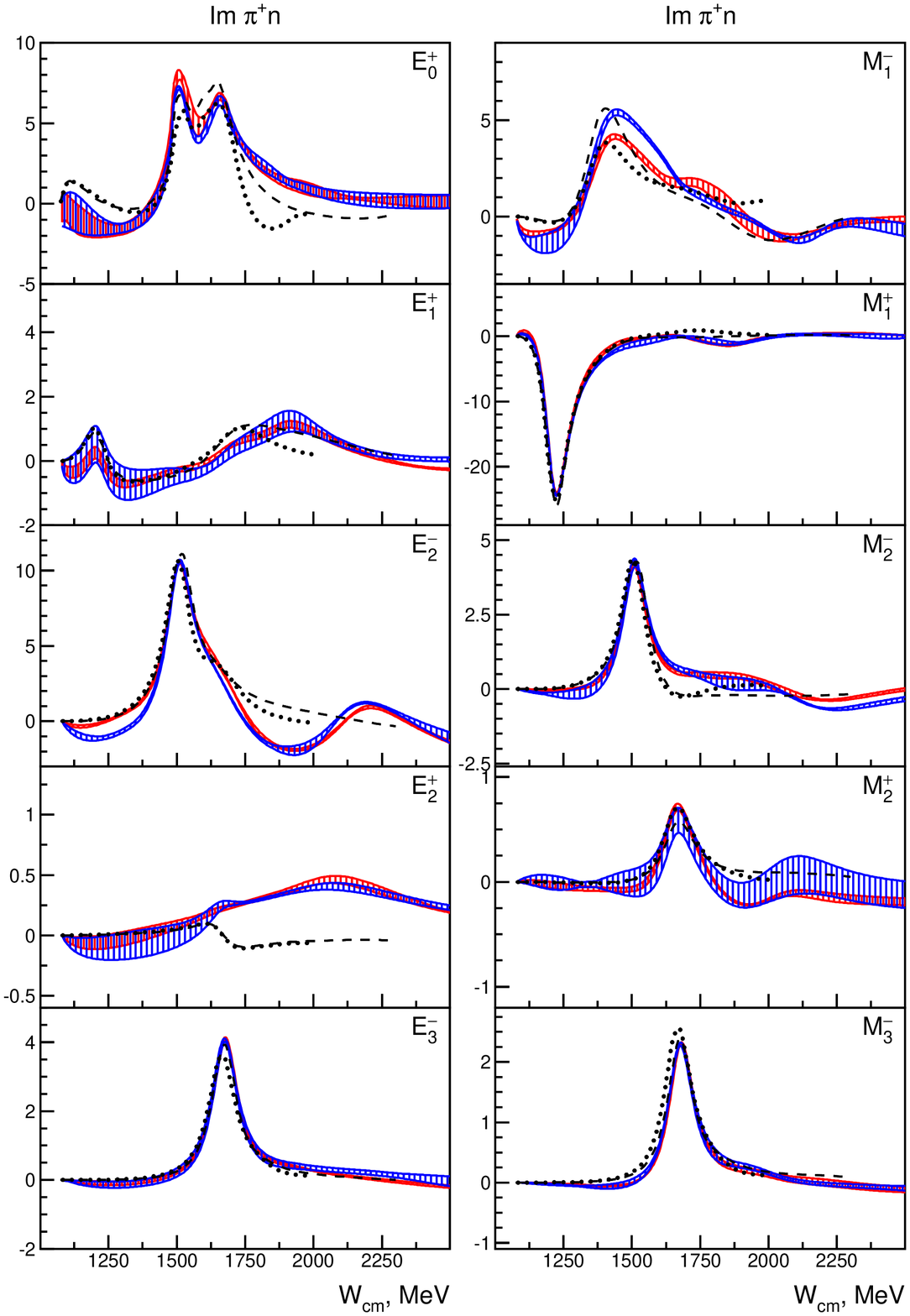}
\end{tabular}
\end{center}
\caption{\label{piN}Multipole decomposition of the $\gamma p\to
p\pi^0$ (top) and $\gamma p\to n\pi^+$ (bottom) transition
amplitudes. The multipoles are given in units of
$10^{-3}/M_{\pi^+}$. The light (red) shaded areas give the range
from a variety of different fits derived from solution BnGa2011-01,
the dark (blue) shaded area from solution BnGa2011-02. The dashed
line represents the SAID fit \cite{SAID}, the dotted line the MAID
\cite{MAID} fit.}
\end{figure*}

Photoproduction multipoles for $\gamma p\to \Lambda K^+$, for the
two isobar contributions to $\gamma p\to\Sigma K$, and for $\gamma
p\to p\eta$ are presented in Figs. \ref{gLambdaK}-\ref{gppeta} and
compared with the amplitudes from KAON-MAID
\cite{Bennhold:1999nd,MAID}. For $\gamma p\to \Lambda K^+$, the two
Bonn-Gatchina solutions show often similar trends in the region up
to 2\,GeV even though the spread of results is still remarkable.
This was unexpected since this reaction is, at present, the best
studied photoproduction reaction for which even data on double
polarization ($C_x, C_x$, $O_x, O_z$) are available, and used in the
Bonn-Gatchina fits.

Here, it has to be stressed that the amplitudes are much more
sensitive to the structure of an amplitude than the position of
singularities in the complex energy plane. Large differences in
these amplitudes may still be compatible with a reasonably good
definition of the masses and widths of resonances.

The KAON-MAID amplitudes show nearly no similarity to our results at
all. However, sizable discrepancies can be seen in some variables
which can be calculated from their amplitudes. Fig.~\ref{piSig}
shows an example. We emphasize that KAON-MAID did not fit those
data. The comparison hence shows that the discrepancies between
KAON-MAID and the Bonn-Gatchina solutions is not too worrisome. The
multipoles for the two isobar contributions to $\gamma p\to\Sigma K$
show an even larger spread, and sizable differences between the two
BnGa solutions, likely due to the absence of data on double
polarization variables. For the reaction $\gamma p\to p\eta$, the
amplitudes exhibit a large spread even in important waves. In this
case, the disagreement between the BnGa and the MAID amplitudes is
very unsatisfactory since basically, the same data are used.
Obviously, care has to be taken when $\eta$ decay modes of
resonances are to be interpreted.

Our multipoles on pion photoproduction were presented already in
\cite{Anisovich:2009zy}. Those were based our 2009 solution. In Fig.
\ref{piN} we give an update based on our present solutions. The
overall consistency between BnGa2009 and BnGa2011, and between
BnGa2011-01 and BnGa2011-02, is good. Also, the error bands are
significantly narrower. The large discrepancies between our
solutions and those of MAID and SAID were, however, unexpected. In
this case, new data from ELSA on the double polarization variable
$G$ provide very useful information \cite{Thiel:2012}. $G$ describes
the correlation between the photon polarization plane and the
scattering plane for protons polarized along direction of the
incoming photon. In the low-energy region ($M<1.65$\,GeV), the data
are compatible with BnGa2011-01 and BnGa2011-02 predictions, while
very sizable discrepancies are seen to the MAID and SAID
predictions. At medium energies ($M\approx 1.65 - 1.80$\,GeV), all
predictions are reasonably close to the data even though
improvements can be expected when the data are included in the fits.

The structure of the $M_1^-$ multipole, exciting the $J^P=1/2^+$
partial wave, is similar in all solutions even though the
quantitative agreement is not satisfying. Other multipoles compare
more favorably, in particular of course the $M_1^+$ multipole.
Significant differences are seen in the $E_2^+$ and $M_2^+$. Yet,
these are the smallest amplitudes (and excite the $N(1675)5/2^-$
resonance).

\section{Summary}
We have presented amplitudes for pion and photo-induced reactions
leading to final states with $\Lambda K$, $\Sigma K$, and $N\eta$.
For completeness, we have added the amplitudes for $\gamma p\to
p\pi^0$ and $\gamma n\to p\pi^+$. The amplitudes are the solutions
of a recent multichannel analysis of the Bonn-Gatchina group
\cite{Anisovich:2011fc}. The number and the properties of
contributing resonances have been reported earlier.

The solutions are divided into two classes, one class in which we
assume that there is one resonance with $J^P=7/2^+$, in the other
class of solutions, it is assumed that there are two of them. Within
each class, there is a variety of solutions, again with different
numbers of assumed resonances, different background
parameterizations, or using different start values of the fit.

In most cases were the BnGa amplitudes can be compared to the
amplitudes from other partial wave analyses, very severe
discrepancies show up. These discrepancies are certainly due to the
lack of polarization data. The gain in reliability in the definition
of amplitudes can be seen when the amplitudes are used to calculate
observable quantities. This can be done using our web page
\cite{BnGa}.

Significant deviations are also observed between the two classes of
solutions BnGa2011-01  and BnGa2011-02 even though both solutions
are compatible with the very large data base made available in the
last years by intense efforts at Bonn, Mainz, Grenoble, and Jlab.
But there is hope for the future: in Bonn, Mainz, and Jlab the
program is ongoing, and experiments with polarized photon beams,
polarized targets, and with measurements of the recoil polarization
are being carried out and were partially completed and are being
analyzed. Hence significant further advances can be expected.

\subsection*{Acknowledgements}
We would like to thank the members of SFB/TR16 for continuous
encouragement. We acknowledge support from the Deutsche
Forschungsgemeinschaft (DFG) within the SFB/ TR16 and from the
Forschungszentrum J\"ulich within the FFE program.

\end{document}